\documentclass[12pt]{amsart}
\usepackage[usenames,dvipsnames]{color}
\usepackage{amsmath,amssymb,amsthm,graphicx,mathrsfs,url,latexsym,enumerate}
\usepackage{color}
\usepackage{a4wide}
\usepackage{ifthen}
\usepackage{verbatim}
\usepackage{fancyhdr}
\usepackage{url}
\usepackage{bbm}
\usepackage[sans]{dsfont}
\usepackage{MnSymbol}

\usepackage{amsbsy}
\usepackage[colorlinks=true,linkcolor=red,citecolor=green]{hyperref}
\usepackage{wrapfig}
\usepackage{tikz}
\usepackage{geometry}
\definecolor{bleu}{RGB}{27,88,145}
\definecolor{mauve}{RGB}{138,20,79}

\renewcommand{\Re}{\operatorname{Re}}

\newcommand{\supp}{\operatorname{supp}}

\newcommand{\diag}{\operatorname{diag}}

\newcommand{\C}{\mathbb C}

\newcommand{\N}{\mathbb N}
\newcommand{\R}{\mathbb R}

\newcommand{\indic}{\operatorname{1\negthinspace l}}

\newcommand{\Ran}{\operatorname{Ran}}
\newcommand{\Ker}{\operatorname{Ker}}

\newcommand{\argmax}{\operatorname{argmax}}
\newcommand{\bsigma}{\boldsymbol{\sigma}}

\def\<{\langle}
\def\>{\rangle}
\newcommand{\bp}{{\it Proof. }}
\newcommand{\ep}{\hfill $\square$\\}

\def\m{\mathbf{m}}

\def\s{\mathbf{s}}

\newcommand{\be}{\begin{equation}}
\newcommand{\ee}{\end{equation}}
\newcommand{\bes}{\begin{equation*}}
\newcommand{\ees}{\end{equation*}}

\numberwithin{equation}{section}
\numberwithin{figure}{section}

\newtheorem{theorem}{Theorem}
\newtheorem{defin}{Definition}[section]
\newtheorem{corollary}[defin]{Corollary}
\newtheorem{lemma}[defin]{Lemma}

\newtheorem{proposition}[defin]{Proposition}
\newtheorem{remark}[defin]{Remark}

\def\ddd{{\mathcal D}}
\def\eee{{\mathcal E}}

 \title[Eyring-Kramers type formulas for PDMP's]{Eyring-Kramers type formulas for some piecewise deterministic Markov processes}
\author[D. Le Peutrec]{Dorian Le Peutrec}
\address{D. Le Peutrec, Institut Denis Poisson, Universit\'e d'Orl\'eans}
\email{dorian.le-peutrec@univ-orleans.fr}

\author[L. Michel]{Laurent Michel}
\address{L. Michel, Institut Math\'ematique de Bordeaux, Universit\'e de Bordeaux}
\email{laurent.michel@math.u-bordeaux.fr}

\author[B. Nectoux]{Boris Nectoux}
\address{Boris Nectoux, Laboratoire de Mathématiques Blaise Pascal, Université Clermont Auvergne}
\email{boris.nectoux@uca.fr}

\begin{document} 
 \maketitle
    \begin{abstract}
    In this work, we give sharp asymptotic equivalents   in the small temperature regime of the smallest eigenvalues of the generator of some  piecewise deterministic Markov processes (including the Zig-Zag process and the Bouncy Particle Sampler process) with refreshment rate  $\alpha$ on the one-dimensional torus $\mathbb T$. These asymptotic  equivalents are usually called Eyring-Kramers type formulas in the literature. 
    %Such a process is   non diffusive and non reversible. 
The case when the refreshment rate $\alpha$ vanishes on $\mathbb T$ is also considered. 
%These asymptotic equivalents provide  a sharp exponential decay rate of the associated semigroup  in $L^2$.   %The proof is original since it shows that these equivalents can be obtained using known results on the asymptotic behavior of the smallest eigenvalue of the  Witten Laplacian,  an elliptic and self-adjoint operator.  
    \medskip

 \end{abstract}

\section{Introduction}\label{sec1}
\subsection{Purpose and motivation}
Piecewise Deterministic Markov Processes~\cite{davis1984piecewise}  (PDMP hereafter) have recently attracted a lot of attention  for their use within the Markov Chain Monte Carlo methodology. Given a potential $U:\mathsf M\to \mathbb R$ on a $d$-dimensional manifold $\mathsf M$ without boundary, such processes are indeed ergodic with respect to the    Gibbs measure 
$$\pi(dx)= \frac{e^{-\frac 2hU(x)}}{\int_{\mathsf  M}e^{-\frac 2hU}} dx, $$
 $dx$ being the Lebesgue measure on the position space $\mathsf M$.
Here, the parameter $h>0$  is proportional to the Boltzmann constant $k_B$ through the relation $h=k_BT$, $T$ being the temperature of the underlying system.     When $h>0$ is fixed, the ergodic properties and the rate of convergence of such processes with or without a refreshment rate $\alpha$, have for instance been studied 
in~\cite{Bzit,BDB,dgm,DMS,roussel,benaim2015qualitative,miclo2013etude,monmarche2014piecewise} (see also references therein). We also refer to \cite{BiVe21} for a spectral analysis of the generator of the  one-dimensional zigzag process and its corresponding semigroup when $h$ is fixed.  \medskip

\noindent
 In many applications in statistical physics where one needs to sample from $\pi$, the constant~$h$ is very small compared to the   energetic barriers of $U$.  
Recently, the long time behavior of the semigroups generated by the generators $\mathsf L_h$ of these processes on $L^2$ have been investigated in~\cite{GuNe20}  when $h\ll 1$,  where it has also  been  proved that in the set   $\{\Re(z)\geq-\epsilon_0 h\}$, $\mathsf L_h$  has exactly $n_0$ eigenvalues      ($n_0$ being the number of local minima of $U$), which are non positive, real,  and exponentially small as $h\to 0$. Such results exhibit a metastable behavior of the PDMP when $h\ll 1$, as it is the case for diffusion processes~\cite{di-gesu-lelievre-le-peutrec-nectoux-17}. In this work, we want to push the analysis of the metastability of the PDMP  further by proving that  each of these $n_0$ eigenvalues   satisfies a so-called Eyring-Kramers type formula when $h\to 0$. Such sharp formulas describe completely the successive timescales involved in the convergence of
the semigroup generated by $-\mathsf L_h$ to $\pi$, as it has been done in \cite{BoGaKl05_01} (see also \cite{HeKlNi04_01})
for elliptic reversible diffusions.

Let us be more precise on our results. We work in the following setting:  $d=1$ (i.e. the dimension of the position space  is equal to $1$) and   $\mathsf M=\mathbb T$,  the one-dimensional torus. In addition, we work with the operator $\mathsf P_h$ defined below in~\eqref{eq:defP}   which is (up to a multiplication by $-h$) unitary equivalent to $\mathsf L_h$, see~\eqref{eq.equiv}. Thus,  all our results  are easily translated in terms of $\mathsf L_h$. 
The purpose of this work is to   compute   sharp asymptotic equivalents of   the $n_0$ smallest  eigenvalues of $\mathsf P_h$ (or equivalently, those of $\mathsf L_h$) in the limit  $h\to 0$, see Theorems~\ref{th:doublewell} and~\ref{th:main}. In particular these asymptotic equivalents allow us to provide a sharp exponential decay rate of the  semigroup  associated to $-\mathsf P_h$ in $L^2$, as $h\ll 1$, see Corollary~\ref{co:1}. These results hold when~\eqref{eq.H1} is satisfied, which implies that  the refreshment rate $\alpha$ does not vanish at critical points of $U$ (i.e. when $\partial_xU=0$), see~\eqref{eq.H1-c}. The case   $\alpha=0$  when $\partial_xU=0$ is  investigated in Section~\ref{sec.alpha=0}, where we show that the smallest eigenvalues of $\mathsf P_h$ satisfy different asymptotic equivalents as $h\to 0$. 
 \medskip
 
  \noindent
  To compute sharp asymptotic equivalents of   the $n_0$ smallest  eigenvalues of $\mathsf P_h$ we proceed as follows.  We first introduce  a suitable change of variables to turn the eigenvalue problem  $(\mathsf P_h-\lambda)u=0$
  % (a non reversible and non diffusive operator, see~\eqref{eq:defP}) 
  into a  nonlinear eigenvalue problem: $(\Delta_{V,h}-\lambda W+\lambda^2) g=0$, where $V=-U$, $\Delta_{V,h}$ is the Witten Laplacian associated with $V$, and $W =2|\partial_x U| +\alpha$  (see~\eqref{eq:defQlambda} and Lemma~\ref{lem:elemQ}). 
Using a Grushin problem and known results on the low-lying spectrum of  $\Delta_{V,h} $, we  prove  that  if $\lambda$ is small enough, the kernel   of $\Delta_{V,h} + \lambda W+ \lambda^2$ is composed of  the singularities of a holomorphic function  $\lambda\mapsto E_{-+}(\lambda)\in \mathbb C^{n_0\times n_0}$ %of the form $\mathsf M_{V,h}-\lambda\mathsf W_h+\mathsf G_h(\lambda)$ 
(see~\eqref{eq:matEmp}). We finally investigate the localization of the singularities of $\lambda\mapsto E_{-+}^{-1}(\lambda)$  to deduce sharp asymptotic equivalents of   the $n_0$ smallest  eigenvalues of $\mathsf P_h$.

\subsection{Setting}\label{sec1.1}
Let $\mathsf E=\{(x,v)\in {\mathbb T}\times \{\pm 1\}\}$ where ${\mathbb T}=\mathbb R/\mathbb Z$ is  the one-dimensional torus. 
Consider the following unbounded operator  $\mathsf P_h$  on $L^2( \mathsf E)$ associated with a smooth  function $U:{\mathbb T}\rightarrow\R$  defined by 
\be\label{eq:defP}
\mathsf P_h=-v\, \mathsf  d_{U,h}+2(v\,\partial_xU)_+(\mathsf I -\mathsf B)+\alpha (\mathsf I-\pi_v)
\ee
where 
$\mathsf  d_{U,h}=h\partial_x+\partial_xU$, $\alpha:{\mathbb T}\rightarrow\R$ is a  $\mathcal C^\infty$ non-negative function (the refreshment rate),    $h>0$ is a parameter proportional to the temperature of the underlying statistical system, $\mathsf I$ is the identity operator, and 
$\mathsf B$ and $\pi_v$ are  defined by
$$
\forall (x,v)\in\mathsf  E,\; \mathsf Bf(x,v)=f(x,-v)
$$
and 
$$
\pi_vf(x,v)=\frac 12(f(x,+1)+f(x,-1))
$$
for all  $f\in L^2(\mathsf E)$. Here and in the following, for $u\in L^2(\mathsf E)$, we denote by  $u_\pm=\max(0,\pm u)$.
\noindent
The operator $\mathsf P_h$ is linked to the   Zig-Zag (or the Bouncy Particle Sampler generator)    process generator   $\mathsf L_h$ where  $\mathsf L _h= v  \partial_x - (\frac 2hv\partial_xU)_+ (\mathsf I-\mathsf B) -\frac 1h \alpha(\mathsf I-\pi_v)$ through the relation 
\begin{equation}\label{eq.equiv}
\mathsf P_h =-  h \, e^{-\frac 1hU}\,  \mathsf L _h  e^{\frac 1hU}.
\end{equation}
We refer to~\cite{bierkens2019zig,BPS1,BPS2} and references therein for more details on these two processes (see also~\cite{faggionato2009non} and \cite[Section 3.1]{durmus2018}). 
%The operator $\mathsf P_h$ is linked to    the Bouncy Particle Sampler  or the  process generator  $\mathsf L^{\mathsf {BPS}}_h$ where  $\mathsf L^{\mathsf {BPS}}_h= v\cdot \pa_x - \lambda_{h,\mathsf{J}} (\mathsf I-\mathsf B) -\frac 1h \lambda_r (\mathsf I-\pi_v)$ introduced in~\cite{BPS1} (see also~\cite{BPS2}, \cite[Section 3.1]{durmus2018},  and Remark~\ref{re.BPS} below)
The space 
$E={\mathbb T}\times\{\pm 1\}$ is endowed with the natural scalar product
\be\label{eqsp1}
\<f,g\>_{L^2(\mathsf E)}=\frac 12\sum_{v=\pm 1}\int_{{\mathbb T}}f(x,v)\overline{g(x,v)}dx.
\ee
 Throughout   this work, we will assume that 
$$\text{$U:{\mathbb T}\to \mathbb R$ is a smooth Morse function}.$$
By definition, this means   that $\partial_x^2U(x)\neq 0$ when $\partial_xU(x)=0$, $x\in {\mathbb T}$. In particular, $U$ has a finite number of critical points on ${\mathbb T}$.
It is proved in  \cite{GuNe20} that $\mathsf P_h$ with domain $D(\mathsf P_h)=\{f\in L^2(\mathsf E),\;v\partial_xf\in L^2(\mathsf E)\}$ is maximal accretive.  
In \cite{GuNe20}, the authors prove the following spectral result on $\mathsf P_h$ in the limit $h\to 0$. 
 \begin{theorem}\label{th:1}($\,${\cite[Theorem 1 and Proposition 13]{GuNe20})}
 Assume that $U$ is a Morse function with $n_0$ local  minimum points. 
 Assume also that   
\begin{equation}
\label{eq.H1}
\min_{\mathbb T}  \big(\, 2\vert \partial_xU\vert +\alpha\,\big)>0.
\end{equation}
  Then, there exist $\epsilon_0>0$  and
 $h_0>0$ such that for all $h\in]0,h_0]$,   $\sigma(\mathsf P_h)\cap\{\Re(z)\leq\epsilon_0 h^2\}$ is made of  $n_0$ real nonnegative eigenvalues $ \lambda_{1,h}\leq \ldots\leq \lambda_{n_0,h}$ (counted with algebraic multiplicity).  Moreover,  %these eigenvalues are real, 
 their algebraic multiplicities   equal their 
geometric multiplicities and there exist $C>0$ and $h_0>0$ such that
% for all $j=1,\ldots,n_0$ and 
for all $h\in (0,h_0]$, 
 $
 %0\leq 
 \lambda_{n_{0},h}\leq e^{-C/h}$. Finally, $\lambda_{1,h}=0$ and has algebraic multiplicity $1$.
 % for $\mathsf P_h$. 
\end{theorem} 
\noindent
Let us mention that  $\dim\Ker  (\mathsf P_h)=1$ holds for all $h>0$ (see \cite{GuNe20}).
In {\cite{GuNe20}, it is assumed that $\min_{\mathbb T}  \alpha>0$, but all the results of \cite{GuNe20} still hold if the less stringent assumption~\eqref{eq.H1} is satisfied.  Indeed,  if~\eqref{eq.H1} holds and $u\in D(\mathsf P_h)$, $\text{Re }\langle \mathsf P_hu,u \>_{L^2(\mathsf E)} \ge  r_m\Vert (\mathsf I-\pi_v)u\Vert^2$, 
where $r_m= \min_{\mathsf E}  \, 2|\partial_xU|+\alpha$. This follows from the following computations. By  \cite[Lemma 5]{GuNe20},
 since $(\mathsf I -\mathsf B)=2(\mathsf I-\pi_v)$,  for all $u\in D(\mathsf P_h)$, one has, denoting by $w= (\mathsf I-\pi_v)u$ (notice that $\vert w(\cdot,1)\vert= \vert w(\cdot,-1)\vert$):  
\begin{align*}
\text{Re }\langle \mathsf P_hu,u \>_{L^2(\mathsf E)}&= \frac 12 \int_{\mathsf E} (2v\,\partial_xU)_+  |(\mathsf I-\mathsf B)u|^2+  \int_{\mathsf E}  \alpha | (\mathsf I-\pi_v)u|^2\\
&=    \int_{\mathsf E} 4(v\,\partial_xU)_+ |w|^2+  \int_{\mathsf E}  \alpha |w|^2\\
&=\frac 12\int_{\mathbb T} 4(\partial_xU)_+ |w|^2+\frac 12\int_{\mathbb T} 4(\partial_xU)_-|w|^2+ \int_{\mathsf E}  \alpha |w|^2 =\int_{\mathsf E}  (2|\partial_xU|+\alpha ) |w|^2.
\end{align*}
  Notice that~\eqref{eq.H1}  implies that $\alpha$ can vanish on $\mathbb T$ but not everywhere since~\eqref{eq.H1}  is equivalent to: 
  \begin{equation}
\label{eq.H1-c}
\text{for any } x\in \mathbb T,  \  \ \partial_x U(x)=0 \ \Rightarrow  \ \alpha(x) >0.
\end{equation} 
 The case when there exists $x\in \mathbb T$ such that  
$\partial_x U(x)=\alpha(x)=0$
% $\alpha(x)=0$ when  $\partial_x U(x)=0$  
is be considered in Section~\ref{sec.alpha=0}. 
 \medskip
 
%and so $\mathsf P_h$ also writes:
 %$$\mathsf P_h=-v\, \mathsf  d_{U,h}+\big[2(v\,\partial_xU)_++\alpha \big](\mathsf I-\pi_v).$$

%\begin{remark}
 %Theorem~\ref{th:1} improves~\cite[Theorem 1]{GuNe20}  as far as the dimension $1$ is concerned.  
 %we do not assume here that $\alpha>r_0$ on $\mathbb S$ for some $r_0>0$. Moreover, the separation of the spectrum of $\mathsf P_h$ is of order $\epsilon_0 h$   in  Theorem~\ref{th:1} and not $\epsilon_0 h^2$ as in~\cite[Theorem 1]{GuNe20}.  
%\end{remark}
\noindent
%Let us recall that the aim  of this work is to compute sharp equivalents of the above   eigenvalues $\{\lambda_1,\ldots,\lambda_{n_0}\}$ as $h\to 0$. 
For our analysis, it will be convenient to introduce the function 
$$V=-U,$$
 which is also a Morse function on $\mathbb T$. We shall denote by  $\mathsf U^{(0)}$ the set of
local  minima of $V$ and by $\mathsf U^{(1)}$ the set of its local maxima. Since we are on the torus, the set $\mathsf U^{(0)}$ and $\mathsf U^{(1)}$ have the same cardinality $n_0$. 

Throughout the paper, we will say that a family of complex numbers $(a_h)_{h>0}$ admits a classical expansion in power of $h^\beta$ (where $\beta>0$) if there exists a sequence 
$(a^k)_{k\geq 0}$ such that, for all $K\geq 0$, one has $a_h=\sum_{k=0}^Ka^k h^{\beta k}+O(h^{\beta(K+1)})$ when $h\rightarrow 0$. In that case, we will denote 
$a_h\sim\sum_{k\geq 0}a^k h^{\beta k}$.

\subsection{Main results}\label{sec1.3}
In this section we give sharp asymptotics of the exponentially small eigenvalues $\lambda_{j,h}$, $j=1,\ldots,n_0$.
Observe that if $n_0=1$, there is only one small eigenvalue by Theorem~\ref{th:1} which is $0$, and there is thus nothing to compute. We then consider the case when $n_0\ge 2$. 
We start with the following theorem which gives the result in the simplified setting of a non-symmetric double well potential $V$   (see Theorem~\ref{th:main}  below for the general case $n_0\ge 2$).
\begin{theorem}\label{th:doublewell} Let $U$ be a Morse function. 
Assume that \eqref{eq.H1} holds and that $\mathsf U^{(0)}$ is made of two elements $\m_1$ and $\m_2$ such that $V(\m_1)<V(\m_2)$.  Assume also that the two elements $\s_1,\s_2$ of 
$\mathsf U^{(1)}$ satisfy $V(\s_1)>V(\s_2)$. Let $\epsilon_0>0$ be as in Theorem~\ref{th:1}.  Then, the second smallest eigenvalue $\lambda_{2,h}$ of $\mathsf P_h$ satisfies  as $h\to 0$: 
$$\lambda_{2,h}=\zeta_h\,h\,e^{-\frac 2h (V(\s_2)-V(\m_2))}, \text{ where   
$\zeta_h\sim \sum_{k\geq 0}h^{\frac k 2}\zeta_k$ and   $
\zeta_0=\frac 1{2\pi}\frac{\sqrt{|V''(\m_2)V''(\s_2)|}}{\alpha(\m_2)}$. }$$
%\item[-] if $\alpha(m_2)= 0$ one has
%$\zeta(\m_2,h)\sim \sqrt{h}\sum_{k\geq 0}h^{\frac k 2}\zeta_{k}(\m_2)$ with 
%and 
%$$
%\zeta_{0}(\m_2)=\frac 12\sqrt{\frac {|V''(\s_2)|}{\pi}}
%$$
%\end{itemize}

%Moreover, in the case where $\alpha$ vanishes identically, the expansion has only integer powers.

%{\bf Ici (et dans le thm 3), verifier que le cas ou $\alpha$ peut s'annuler est traitable. Il  me semble que dans \cite{GuNe20}, on suppose $\alpha(x)>\alpha_0>0$}
\end{theorem}
%\noindent
% \textcolor{blue}{...il y a avait un "assume for simplicity $V(\s_1)=V(\s_2)$" dans la version d'avant} 

The situation where $V(\m_1)=V(\m_2)$ and/or $V(\s_1)=V (\s_2)$ could be handled easily by constructing adapted quasimodes in the spirit of \cite{Mi19}. Here, we decided to state our result in the above simplified setting in order to lighten the formulas.

%\textcolor{blue}{Ici, il faudrait mettre un petit commentaire, peut etre comparer avec le thm de Pierre Monmarch\'e}
%\subsection{Main result}\label{sec1.2}

Let us now  state our result in the  general setting $n_0\geq 2$.
To this end,
%In order to state our main result in this case, 
we need to label the local minima and maxima  of $V$ in a suitable way. The following construction is inspired  from 
\cite{HeHiSj11_01} (see also  \cite{Mi19} and \cite{LePMi20}).
In order to simplify,  we assume from now that $V$ uniquely attains its maximum at the point  $\s_{max}\in\mathsf U^{(1)}$, i.e. that
\begin{equation}\label{eq.H01}
\argmax_{\mathbb T} V= \{\s_{max}\}.
\end{equation}
Then, set  
$\underline{\mathsf U}^{(1)}=\mathsf U^{(1)}\setminus \{\s_{max}\}$. Since $n_0\ge 2$,   $\underline{\mathsf U}^{(1)}\neq\emptyset$. 
We denote the elements of $V(\underline{\mathsf U}^{(1)})$ by $\sigma_2>\sigma_3>\ldots>\sigma_N$, where $N\geq 2$.  For convenience, we also introduce a fictive infinite saddle value $\sigma_1=+\infty$ and we denote $\Sigma=\{\sigma_1,\ldots,\sigma_N\}$.
Starting from $\sigma_1$, we will recursively associate to each $\sigma_i$ a finite family of local minima 
$(\m_{i,j})_j$ and a finite family $(\mathsf C_{i,j})_j$ of connected components of $\{V<\sigma_i\}$ in the following way

\begin{itemize}
\item[$\star$] Let $X_{\sigma_{1}} = \{ x \in {\mathbb T} ; \ V ( x ) < \sigma_{1} = + \infty \} = {\mathbb T}$. We let $\m_{1 , 1}$ be any global minimum of $V$
 (not necessarily unique) and $\mathsf C_{1 , 1} = {\mathbb T}$. In the following, we will denote $\underline{\m} = \m_{1 , 1}$.

\item[$\star$] Next we consider $X_{\sigma_{2}} = \{ x \in {\mathbb T} ; \ V ( x ) < \sigma_{2} \}$. This is the union of its finitely many connected components. Exactly one of these components contains $\m_{1 , 1}$ and the other components are denoted by $\mathsf C_{2 , 1} , \ldots , \mathsf C_{2 , N_{2}}$. In each component $\mathsf C_{2 , j}$, we pick up a point $\m_{2 , j}$ which is a global minimum of $V_{\vert \mathsf C_{2 , j}}$.

\item[$\star$] Suppose now that the families $( \m_{k , j} )_{j}$ and $( \mathsf C_{k , j} )_{j}$ have been constructed until rank $k = i - 1$. The set $X_{\sigma_{i}} = \{ x \in {\mathbb T} ; \ V ( x ) < \sigma_{i} \}$ has again finitely many connected components and we label $\mathsf C_{i , j}$, $j = 1 , \ldots , N_{i}$ those of these components which do not contain any $\m_{k , \ell}$ with $k < i$. In each $\mathsf C_{i , j}$ we pick up a point $\m_{i , j}$ which is a global minimum of $V_{\vert \mathsf C_{i , j}}$. 
%Observe that for all $i \geq 2$, the components $\partial \mathsf C_{i , j}$ contains at least one element of $\underline{\mathsf U}^{(1)}$.
\end{itemize}
We run the procedure until all the minima have been labeled. 
\ 
\begin{remark}
Since we work on ${\mathbb T}$, using the terminology of \cite{HeHiSj11_01},  every maximum point $\s\in\underline{\mathsf U}^{(1)}$ is a separating saddle points (ssp) and $\s_{max}$ is 
not a ssp. In the case where $V$ attains its global maximum at several  distinct points $\s_{max,1},\ldots,\s_{max,k}$, every maximum point is separating and 
the situation can be handled easily by a modification of the construction above.
However, this would lead to a slightly more complicated presentation of the results that we prefer to avoid in this work.
\end{remark}
\noindent
We now recall some constructions of \cite{Mi19} and \cite{LePMi20} that will be useful in the sequel. 
Throughout we denote $\underline{\mathsf U}^{(0)}=\mathsf U^{(0)}\setminus\{\underline{\m}\}$, $\s_{1}$ is a fictive saddle point such that $V ( \s_{1} ) = \sigma_{1} = + \infty$. For any set $A$, ${\mathcal P} ( A )$ denotes the power set of $A$. From the above labelling we define   two mappings
\begin{equation*}
\mathsf C : {\mathsf U}^{( 0 )} \to {\mathcal P} ( \R^{d} ) \qquad \text{and} \qquad {\bf j} : {\mathsf U}^{( 0 )} \to {\mathcal P} ( \underline{\mathsf U}^{(1)} \cup \{ \s_{1} \} ) ,
\end{equation*}
 as follows: for every $i \in \{ 1 , \dots , N \}$ and $j \in \{ 1 , \dots , N_{i} \}$,
\begin{equation} \label{a25}
\mathsf C ( \m_{i , j} ) : = \mathsf C_{i , j} ,
\end{equation}
and
\begin{equation} \label{a26}
{\bf j} ( \underline{\m} ) : = \{ \s_{1} \} \qquad \text{and} \qquad {\bf j} ( \m_{i , j} ) : = \partial \mathsf C_{i , j} \cap \underline{ \mathsf U}^{( 1 )} \text{ for } i \geq 2 .
\end{equation}
In particular, we have $\mathsf C( \underline{\m} ) = {\mathbb T}$ and
 for all
$i, j \in \{ 1 , \dots , N \} $,  one has $\emptyset \neq {\bf j} ( \m_{i , j} ) \subset
\{ V = \sigma_{i} \} $.
We then define the mappings
\begin{equation*}
\bsigma :  {\mathsf U}^{(0)} \rightarrow \Sigma \qquad \text{and} \qquad S :  {\mathsf U}^{(0)} \rightarrow ( 0 , + \infty ] ,
\end{equation*}
by
\begin{equation} \label{a27}
\forall \m \in \mathsf U^{( 0 )} , \qquad \bsigma ( \m ) : = V ( {\bf j} ( \m ) ) \qquad \text{and} \qquad S ( \m ) : = \bsigma ( \m ) - V ( \m ) ,
\end{equation}
where, with a slight abuse of notation, we have identified the set $V ( {\bf j} ( \m ) )$ with its unique element. Note that $S ( \m ) = + \infty$ if and only if $\m = \underline{\m}$.
%Given $\m \in \ulu^{( 0 )}$, one has $\bsigma ( \m ) = \sigma_{i}$ for some $i \geq 2$. Moreover, since $\sigma_{i - 1} > \sigma_{i}$, there exists a unique connected component of  $\{ f < \sigma_{i - 1} \}$ that contains $\m$ (observe that this component is not necessarily critical). We denote that component by $E_{-} ( \m )$, and by
%\begin{equation} \label{b30}
%E_{-} : \ulu^{( 0 )} \rightarrow {\mathcal P}( \R^{d} ) ,
%\end{equation}
%the corresponding application. It follows from  Remark 2.2 of \cite{Mi19} that, for any $\m \in \ulu^{( 0 )}$, there exists a unique $\widehat{\m} \in E_{-} ( \m ) \cap \mathsf U^{(0)}$, denoted by $\widehat{\m} ( \m )$, such that $\bsigma ( \widehat{\m} ) > \bsigma ( \m )$.  In particular,
%\begin{equation} \label{b31}
%\forall \m \in \ulu^{( 0 )} , \qquad f ( \widehat{\m} ( \m ) ) \leq f ( \m ) ,
%\end{equation}
%and we denote  by $\widehat{E} ( \m )$ the connected component of $\{ f < \sigma ( \m ) \}$ containing $\widehat{\m} ( \m )$. It holds additionally $\widehat{E} ( \m ) \subset E_{-} ( \m )$ and we can easily see that $\widehat{E} ( \m )$ is always a critical component (see Definition \ref{a23}). We denote by $\widehat{E} : \ulu^{( 0 )} \rightarrow \Cr$ and $\widehat{\m} : \ulu^{( 0 )} \rightarrow \mathsf U^{(0)}$ the corresponding applications.
%
\noindent
With the above notations, our last assumption is the following:
\begin{equation} \label{eq.H2}
\begin{aligned}
&\star\text{ Equation~\eqref{eq.H01} is satisfied}.\\
&\star\text{ For any } \m \in \mathsf U^{( 0 )} , \ \m \text{ is the unique global minimum of } V_{\vert \mathsf C ( \m )}.  \\
&\star\text{ For all } \m^{\prime} \in \mathsf U^{( 0 )} \setminus \{ \m \} , \ {\bf j} ( \m ) \cap {\bf j} ( \m^{\prime} ) = \emptyset .
\\
&\star \text{ The map $S$: ${\mathsf U}^{(0)} \rightarrow ( 0 , + \infty ]$ is injective.}
\end{aligned}
\end{equation}
In particular, \eqref{eq.H2} implies that $V$ uniquely attains its global minimum on $\mathbb T$ at $\underline \m \in \mathsf U^{( 0 )}$. 
 In the following, when \eqref{eq.H2} holds, we label 
the local minima $\m_{1} , \ldots , \m_{n_{0}}$ of $V$ such that $( S ( \m_{j} ) )_{j \in \{ 1 , \dots , n_{0} \}}$ is decreasing (see \eqref{a27}), that is 
\begin{equation} \label{eq.order}
 \text {  for all } j \in \{ 1, \dots , n_{0}-1 \} , \ S ( \m_{j + 1} ) < S ( \m_{j} )  \text{ and } S({\m_1})=+\infty \ \text{(i.e. $\m_1=\underline{\m}$).}
\end{equation}
 \begin{remark}\label{re.REDW}
 Notice that, in the geometrical setting of Theorem~\ref{th:doublewell}, one has  by construction of $S$   (see~\eqref{a27})  and $\bf j$ (see~\eqref{a26}), 
$$S(\m_2)=V(\s_2)-V(\m_2) \text{ and } \bf j(\m_2)=\{s_2\}.$$
 \end{remark}
 \noindent
The main result of this work is the following. 
\begin{theorem}\label{th:main} Let $V=-U$ be a Morse function.  
Assume that \eqref{eq.H1} and  \eqref{eq.H2} are satisfied. Let $\epsilon_0>0$ be given by 
Theorem \ref{th:1}. Then, there exists $h_0>0$ such that, for all 
$h\in]0,h_0]$,  the  $n_0$ eigenvalues 
%real nonnegative
$\lambda_{1,h}\leq\ldots\leq\lambda_{n_0,h}$
%$\{\lambda_{1,h},\ldots,\lambda_{n_0,h}\}$  
of $\mathsf P_h$ in $\{\Re(z)\leq\epsilon_0 h^2\}$ (counted with algebraic multiplicity)
% (orderer such that $\lambda_{j,h}\le \lambda_{j+1,h}$ for all $j=1,\ldots,n_0-1$)  
satisfy: $\lambda_{1,h}=0$ and, for all $j\in \{2,\ldots,n_0\}$,  
\begin{equation} \label{e12}
\lambda_{j,h} =  \zeta_h ( \m_j)\,h\, e^{- \frac 2h S ( \m_j ) }  ,
\end{equation}
where 
$S : \mathsf U^{( 0 )} \rightarrow ( 0 , + \infty ]$ is defined in \eqref{a27} (see also \eqref{eq.order})
and $\zeta_h$ admits a classical expansion
 $ \zeta_h(\m_j)\sim \sum_{k\geq 0}h^{\frac k 2} \zeta_k(\m_j)$ with
\begin{equation} \label{e13}
 \zeta_0 ( \m_j ) = \frac{1}{2 \pi}\sum_{\s \in {\bf j} ( \m_j )} \frac{\sqrt{|V'' ( \m_j )V'' ( \s )|}}{\alpha(\m_j)},  
\end{equation}
where ${\bf j} : \mathsf U^{( 0 )} \to \mathcal P ( \underline{\mathsf U}^{( 1 )} \cup \{ \s_{1} \} )$ is defined in \eqref{a26}.
%\item[-] if $\alpha(\m)= 0$, then $ \zeta(m,h)\sim \sqrt h\sum_{k\geq 0}h^{\frac k 2} \zeta_k(\m)$ with 
%\begin{equation} \label{e13}
% \zeta_0 ( \m ) = \frac{1}{2 }\sum_{\s \in {\bf j} ( \m )}\sqrt{\frac{ |V'' ( \s )|}{\pi}}  ,
%\end{equation}
%\end{itemize} 
 
\end{theorem}
\noindent
Notice that, according to Theorem \ref{th:main},
$\lambda_{j,h}$ is a simple eigenvalue of $\mathsf P_h$ 
 for all $j\in \{1,\ldots,n_0\}$ and $h$ small enough  (that is   $\dim\big(\Ker (\mathsf P_h-\lambda_{j,h})^{m}\big)=1$ for every $m\in\N^{*}$). 
\begin{remark}
The assumptions~\eqref{eq.H01} and  \big($S$: ${\mathsf U}^{(0)} \rightarrow ( 0 , + \infty ]$   is injective\big) in \eqref{eq.H2} 
 are generic. They can be relaxed following the procedure of  \cite{LePMi20}. The whole assumption \eqref{eq.H2} could also be relaxed by following 
 the strategy of \cite{Mi19}.
 \end{remark}
\noindent
Let us recall that by the Hille-Yosida Theorem, $-\mathsf P_h$ generates a strongly continuous contraction semigroup $(e^{-t\mathsf P_h})_{t\ge 0}$ on $L^2({\mathsf E})$. 
Let us recall that under the assumptions of Theorem~\ref{th:main},
according to~\cite[Theorem 2]{GuNe20}, 
we have, for some $c>0$ and every $h>0$ small enough:
$$
e^{-t\mathsf P_h} = \sum_{j=1}^{{n_{0}}} e^{-t\lambda_{j,h}} \Pi_{j,h}  + O(e^{-c\,t\,h^{2}})\ \ \ \text{in}\ \mathcal L(L^{2}(\mathsf E)),
$$
where, for $j=1,\dots,n_{0}$, $\Pi_{j,h}$ is the spectral projector associated with the eigenvalue $\lambda_{j,h}$
of $\mathsf P_h$, and $\Pi_{j,h}=O(1)$ in $\mathcal L(L^{2}(\mathsf E))$.
Using in addition Theorem~\ref{th:main}, we get sharp asymptotic equivalents
of
 the different timescales $ 1/\lambda_{j,h} $ involved in the return to  equilibrium.
This leads in particular to the following accurate exponential decay rate in $L^2(\mathsf E)$  of the semigroup $(e^{-t\mathsf P_h})_{t\ge 0}$ as $h\ll 1$.

\begin{corollary}\label{co:1}  Let $U$ be a Morse function.  
Assume that \eqref{eq.H1} and  \eqref{eq.H2} are satisfied.   
Then, there exist  $C>0$ and $h_0>0$  such that, for all $h\in (0,h_0)$,  it holds for all $t\ge 0$: 
  $$\big \Vert e^{-t\mathsf P_h}-  \Pi_{1,h} \big \Vert_{\mathcal L(L^2(\mathsf E))}\le    C   e^{- t\lambda_{2,h}},$$
  where, as $h\to 0$,  $\lambda_{2,h}=\zeta_h ( \m_2)\,h\, e^{- \frac{2 S ( \m_2 )}{h}}$ with $\zeta_h ( \m_2)\sim \sum_{k\geq 0}h^{\frac k 2} \zeta_k(\m_2)$ and     $\zeta_0(\m_2)$ given by~\eqref{e13}, 
 % (with $j=2$ there), 
  and where $\Pi_{1,h}$ is the $L^2(\mathsf E)$ orthogonal projection on ${\rm Span } (e^{-  U/h}\mathsf 1_{\{\pm 1\}})$, where $\mathsf 1_{\{\pm 1\}}$ is the constant function on $\{\pm 1\}$ which equals $1$.   
\end{corollary}
%More precise metastable transitions in the spirit of
%Corollary 1.6 of \cite{BoLePMi22} could also be proven. 

\subsection{Extension of the results to the case when~\eqref{eq.H1} is not satisfied}\label{sec.alpha=0}
In this section, we assume that 
  \begin{equation}
\label{eq.H1-alpha=0}
\text{for any } x\in \mathbb T,  \  \ \partial_x U(x)=0 \ \Rightarrow  \ \alpha(x) =0,
\end{equation}
and we give asymptotic equivalents of the $n_0$ first eigenvalues of $
\mathsf P_h$ when~\eqref{eq.H1-alpha=0}   holds (and then, in particular, when there is no refreshment at all, i.e. when $\alpha=0$).
%%%%

\begin{theorem}\label{th:alpha} Let $U$ be a Morse function.  
Assume that  \eqref{eq.H1-alpha=0} and \eqref{eq.H2} are satisfied. Then, for any $c>0$, there exists $h_0>0$ such that, for all $h\in (0,h_0]$, 
$\sigma(\mathsf P_h)\cap D(0,ch^2)$
is made of  $n_0$ real   eigenvalues $\lambda_{1,h}\leq\ldots\leq\lambda_{n_0,h}$ (counted with geometric multiplicity), which 
all have geometric multiplicity one.
%are all simple. 
%In addition, for all $j=\{1,\ldots,n_0\}$, $\lambda_{j,h}\in \mathbb R$. 
% Finally, 
In addition, $\lambda_{1,h}=0$ and, for all  $j\in \{2,\ldots,n_0\}$,  
 \begin{equation} \label{e-alpha}
 \lambda_{j,h} =  \zeta_h ( \m_j)\,\sqrt{h}\, e^{- \frac 2h S ( \m_j ) }  ,
 \end{equation}
 where $\zeta_h$ admits a classical expansion 
$ \zeta_h(\m_j)\sim \sum_{k\geq 0}h^{\frac k 2} \zeta_k(\m_j)$ with 
 \begin{equation} \label{e-alpha2}
  \zeta_0 ( \m_j ) =  \frac 14\sum_{\s \in {\bf j} ( \m_{j} )}\sqrt{\frac{ |V'' ( \s )|}{\pi}} .
 \end{equation}
  \end{theorem}
%\begin{remark}
Compared with Theorem \ref{th:main}, we cannot exclude the existence of other eigenvalues in the strip $0<\Re(z)<c h^2$ with large imaginary part.
The reason for this is that Theorem \ref{th:1} does not apply when $\alpha$ and $\partial_x U$ vanish simultaneously.

The prefactor~\eqref{e-alpha2}  is consistent with the one obtained in \cite[Theorem 1.1]{monmarche2014piecewise} for the expected hitting time  of a $1$-dimensional PDMP with no refreshment. Indeed, 
when $\alpha=0$ and 
$V$, and thus $U=-V$,  is a non-symmetric double well potential  (i.e. when $\mathsf U^{(0)}=\{\m_1,\m_2\}$ with $V(\m_1)<V(\m_2)$ and 
$\mathsf U^{(1)}=\{\s_1,\s_2\}$
 with $V(\s_1)>V(\s_2)$),  Theorem~\ref{th:alpha}  and \cite[Theorem 1.1]{monmarche2014piecewise} imply that $h \,\lambda_{2,h}\, \mathbb E[\tau]=1+o(1)$ as $h\to 0$, where $\tau$ is the first   time  the   process $(X_t,Y_t)$  with generator $\mathsf L_h$ on $\mathbb T\times \{\pm 1\}$ hits $\{\m_2\}$ when it starts at  $(\s_2 ,-1)$. Let us also recall here that $h\, \lambda_{2,h}$ is the first nonzero eigenvalue of $-\mathsf L_h$.  
 
 Of course, the situation where the refreshing function $\alpha$ vanishes at some critical points $x\in \mathbb T$ of $U$ but not 
 at
 all could easily be handled and would lead, for each eigenvalue $\lambda_{j,h}$,
 either to the formula given in Theorem~\ref{th:main}
 or to the formula given in Theorem~\ref{th:alpha}, depending on whether $\alpha(\m_{j})>0$
 or  $\alpha(\m_{j})=0$.

 %\end{remark}

\section{Reduction to a finite dimensional  problem} 
\noindent
In this section, we prove that, if $\lambda$ is small enough,  we can reduce the infinite dimensional problem $(\mathsf P_h-\lambda_h)u=0$ into a finite dimensional nonlinear eigenvalue  problem. 
\subsection{A suitable change of variables}
Let us   introduce some notation. For $\lambda\in\C$ and $r>0$ we denote 
$D(\lambda,r)=\{z\in\C,\,|z-\lambda|<r\}$.  For two families of numbers $a = ( a_{h})_{h>0}$ and $b = ( b_{h})_{h>0}$, we say that 
$a \in \eee_{cl} ( b )$ if there exists a family $c=( c_{h})_{h>0 }$ such that $a_{h} = b_{h} c_{h}$ and $c$ admits a classical expansion $c_{h} \sim \sum_{j \geq 0} c_{j} h^{j}$ with $c_{0} = 1$ as $h\to 0$. In all this work, $C>0$ and $c>0$ are constant which are independent of $h$ and which can change from one occurence to another.

Recall that  
$\mathsf E={\mathbb T}\times\{\pm 1\}$ is endowed with the natural scalar product~\eqref{eqsp1}.  Let $\mathsf F=L^2({\mathbb T})\times L^2({\mathbb T})$ and let 
$\<,\>_{\mathsf F}$ be the Hilbertian structure induced by the isomorphism
\be
\begin{split}
\Omega_1:L^2({\mathbb T}\times&\{\pm1\})\rightarrow L^2({\mathbb T})\times L^2({\mathbb T}) \\
&f\longmapsto
\left(\begin{array}{c}
f(.,+1)\\
f(.,-1)
\end{array}
\right).
\end{split}
\ee
Then, $\Omega_1$ is unitary from $(\mathsf E,\<,\>_{L^2(\mathsf E)})$ onto $(\mathsf F,\<,\>_{\mathsf F})$, and 
$$\Omega_1D(\mathsf P_h)= H^1({\mathbb T})\times H^1({\mathbb T}).$$
Direct computations show that 
$$
\Omega_1\, v\, \mathsf d_{U,h}\,\Omega_1^{-1}
=\left(\begin{array}{cc}
\mathsf d_{U,h}&0\\
0&-\mathsf d_{U,h}
\end{array}
\right)%\binom{f_+}{f_-}
,\;\;\;
\Omega_1\, (v\, \partial_xU)_+\,\Omega_1^{-1}=\left(\begin{array}{cc}
(\partial_xU)_+&0\\
0&(\partial_xU)_-
\end{array}
\right)
$$
$$
\Omega_1 \mathsf B\Omega_1^{-1}=\left(\begin{array}{cc}
0&1\\
1&0
\end{array}
\right)
,\;\;\text{
and }\;
\Omega_1 \pi_v\Omega_1^{-1}=\frac 12\left(\begin{array}{cc}
1&1\\
1&1
\end{array}
\right).
$$
Combining these identities with \eqref{eq:defP}, we get
\be\label{eq:Pconj1}
\Omega_1 \mathsf  P_h\Omega_1^{-1}=
\left(\begin{array}{cc}
-\mathsf d_{U,h}&0\\
0&\mathsf d_{U,h}
\end{array}
\right)
+2\left(\begin{array}{cc}
(\partial_xU)_+&-(\partial_xU)_+\\
-(\partial_xU)_-&(\partial_xU)_-
\end{array}
\right)
+\frac \alpha 2\left(\begin{array}{cc}
1&-1\\
-1&1
\end{array}
\right).
\ee
We now change the variables in $\mathsf F$ and consider the unitary transformation $\Omega_2:\mathsf F\rightarrow \mathsf F$ defined by $$\Omega_2(f,g)=\frac 1{\sqrt 2}(f+g,g-f).$$
 Consider the two vectors of $\R^2$ given by 
$e_1=\binom{1}{1}$ and $e_2=\binom{-1}{1}$. Then, one has:
\be\label{eq:Pconj2}
\Omega_1 \mathsf P_h\Omega_1^{-1}(f e_1)= \mathsf d_{U,h} f e_2.
\ee
On the other hand,
\begin{equation*}
\begin{split}
2\left(\begin{array}{cc}
(\partial_xU)_+&-(\partial_xU)_+\\
-(\partial_xU)_-&(\partial_xU)_-
\end{array}
\right)&\binom{-f}{f}=4f\binom{-(\partial_xU)_+}{(\partial_xU)_-}\\
&=2((\partial_xU)_--(\partial_xU)_+)\binom{f}{f}+2((\partial_xU)_-+(\partial_xU)_+)\binom{-f}{f}\\
&=-2\partial_xU f e_1+2|\partial_xU| f e_2. 
\end{split}
\end{equation*}
It follows that:
\be\label{eq:Pconj3}
\begin{split}
\Omega_1 \mathsf P_h\Omega_1^{-1}(f e_2)&=\mathsf d_{U,h} fe_1- 2\partial_xU f e_1+(\alpha+2|\partial_xU|) fe_2\\
&=\mathsf d_{-U,h} fe_1+(\alpha+2|\partial_xU|) fe_2
\end{split}
\ee
Combining \eqref{eq:Pconj2} and \eqref{eq:Pconj3}, we get
$$
\Omega_2\Omega_1 \mathsf P_h\Omega_1^{-1}\Omega_2^{-1}=
\left(\begin{array}{cc}
0&\mathsf d_{-U,h}\\
\mathsf d_{U,h}&\alpha+2|\partial_xU|
\end{array}
\right).
$$
 Set $\mathsf Q_h:=\Omega_2\Omega_1 \mathsf P_h\Omega_1^{-1}\Omega_2^{-1}$ with domain $D(\mathsf Q_h)=H^1({\mathbb T})\times H^1({\mathbb T})$ on $L^2({\mathbb T})^2$, i.e.   
\be\label{eq:defQ}
\mathsf Q_h=\left(\begin{array}{cc}
0&\mathsf d_{-U,h}\\
\mathsf d_{U,h}&W
\end{array}\right)=\left(\begin{array}{cc}
0&\mathsf d_{V,h}\\
-\mathsf d_{V,h}^*&W
\end{array}\right)
\ee
 where we recall that $V=-U$ and where we denote    
$$  W(x):=\alpha(x)+2|\partial_xU(x)|=\alpha(x)+2|\partial_xV(x)|, \ x\in \mathbb T.$$
Since the transformation $\Omega_2\Omega_1$ is unitary from $\mathsf E$ onto $\mathsf F$, 
$$\sigma(\mathsf P_h)=\sigma(\mathsf Q_h).$$   
We will in the following study the spectrum of $\mathsf Q_h$ to prove Theorems~\ref{th:doublewell} and~\ref{th:main}. 
Introduce  the semiclassical Witten Laplacian  $\Delta_{V,h}$ associated with $V$ on ${\mathbb T}$, that is  
$$\Delta_{V,h}=\mathsf d_{V,h}^*\mathsf d_{V,h}=-h^2\Delta+|\partial_xV(x)|^2-h\partial_x^2V(x),\ \text{ with domain $D(\Delta_{V,h})=H^2({\mathbb T})$}.$$
For $\lambda\in \mathbb C$,  we finally define
\be\label{eq:defQlambda}
\mathsf  T_h(\lambda)=\Delta_{V,h}-\lambda W+\lambda^2 \text{ with domain } D(\Delta_{V,h}). 
\ee
The following result is the key point of our analysis: it establishes an equivalence between    the spectrum of $\mathsf Q_h$ and the kernels of the operators $\mathsf  T_h(\lambda)$, $\lambda\in \mathbb C$.
\begin{lemma}\label{lem:elemQ}
The operator $\mathsf Q_h$ with domain $D(\mathsf Q_h)=H^1({\mathbb T})\times H^1({\mathbb T})$ is closed and has compact resolvent.
 In particular, it has only discrete spectrum. 
 Moreover, $\lambda=0$ is a simple eigenvalue of $\mathsf Q_h$ and 
 $\dim(\Ker \mathsf T_h(0))=\dim(\Ker \mathsf Q_h)=1$. Besides, for every  $\lambda\in \mathbb C\setminus\{0\}$, 
 the application 
 \be\label{eq:defPsi}
 \begin{array}{ccl}
 \Psi:&\Ker(\mathsf T_h(\lambda))&\longrightarrow\  \Ker(\mathsf Q_h-\lambda)\\
&% \phantom{*****}
g&\longmapsto \ 
\left(
\begin{array}{c}
\frac 1\lambda \mathsf d_{V,h} g\\
g
\end{array}
\right)
 \end{array}
 \ee
 is a linear isomorphism.
 Eventually, any   $\mu\in \mathbb C$ is a singularity of $\lambda\mapsto \mathsf T_h(\lambda)^{-1}$ if and only if  there exists $g\in H^2(\mathbb T)$, $g\neq 0$, such that $\mathsf T_h(\mu)g=0$. 
\end{lemma}
\noindent
\bp
The operator $\mathsf Q_h$ is closed   on $L^2({\mathbb T})^2$ since it is a bounded perturbation of the closed operator  $(0,h\partial_x;h\partial_x,0)$. 
The resolvent of $\mathsf Q_h$  is compact since the injection $H^1({\mathbb T})\subset L^2({\mathbb T})$ is compact. 
Since $\mathsf Q_h$ is unitarily equivalent to $\mathsf P_h$ (see \eqref{eq:defQ}), it follows from Theorem~\ref{th:1} that 
$\lambda=0$ is a simple eigenvalue of $\mathsf Q_h$.
%$\dim(\Ker Q_h)=1$. 
Moreover,
since $\mathsf T_h(0)=\Delta_{V,h}$, one gets
$\Ker \mathsf T_h(0)=\C e^{-\frac Vh}$
and then $\dim(\Ker \mathsf T_h(0))=1$.

Let us now consider  $\lambda\in\C\setminus\{0\}$. For any $g\in \Ker( \mathsf T_h(\lambda))$, one has $g\in H^2({\mathbb T})$ and thus 
$(\frac 1\lambda  \mathsf d_{V,h} g,g)\in D(\mathsf Q_h)$. 
It follows moreover from 
$\mathsf T_h(\lambda)g=0$ that
$$
\mathsf Q_h\Psi(g)=
\left(\begin{array}{cc}
0&\mathsf d_{V,h}\\
-\mathsf d_{V,h}^*&W
\end{array}\right)
\left(
\begin{array}{c}
\frac 1\lambda \mathsf d_{V,h} g\\
g
\end{array}
\right)
=
\left(
\begin{array}{c}
\mathsf d_{V,h} g\\
(-\frac 1\lambda \Delta_{V,h}+W)g
\end{array}
\right)=\lambda\Psi(g).
$$
 This proves that $\Psi$ is well defined. The linearity and the injectivity of $\Psi$ are obvious.
To prove its surjectivity, consider $u=\binom{f}{g}\in D(\mathsf Q_h)$  such that  
$\mathsf Q_hu=\lambda u$, i.e. such that
\begin{equation*}
\left\{\begin{split}
\mathsf d_{V,h} g&=\lambda f,\\
-\mathsf d_{V,h}^* f+Wg&=\lambda g.
\end{split}
\right.
\end{equation*}
It follows from the first equation that $g\in H^2({\mathbb T})$ and, since $\lambda\neq 0$,
that $u=\binom{\frac 1\lambda \mathsf  d_{V,h}g}{g}$. 
Moreover, applying $\mathsf d_{V,h}^*$ to the second equation leads to
\begin{equation*}
\label{eq:vpPhsyst}
\mathsf \Delta_{V,h}g
=\lambda \mathsf d_{V,h}^* f
%\mathsf d_{V,h}^*\mathsf d_{V,h} g
=\lambda W g-\lambda^2 g.
\end{equation*}
%where the first equation is   in the sense of distribution. 
Consequently,   $g\in D(\Delta_{V,h})$  and
%and \eqref{eq:vpPhsyst} writes   
$\mathsf  T_h(\lambda) g
%= \Delta_{V,h}g-\lambda Wg+\lambda^2g
 =0$, which proves the surjectivity.

 It remains to prove the last statement of Lemma~\ref{lem:elemQ}. 
   To this end, let us  consider $\mu\in \mathbb C$. It is  a singularity of $\lambda\mapsto \mathsf T_h(\lambda)^{-1}$
  if and only if  $\mathsf T_h(\mu):H^2({\mathbb T})\to L^2({\mathbb T})$ is not invertible. Furthermore, since $\Delta_{V,h}+1:H^2({\mathbb T})\to L^2({\mathbb T})$ is invertible,  
   $$\mathsf T_h(\lambda)= \big[1- (1+\lambda W-\lambda^2) (\Delta_{V,h}+1)^{-1}\big](\Delta_{V,h}+1)$$
   is not invertible if and only if $1- \mathsf B(\lambda):L^2({\mathbb T})\to L^2({\mathbb T})$ is not invertible, where $\mathsf B(\lambda)=   (1+\lambda W-\lambda^2)(\Delta_{V,h}+1)^{-1}$. 
 Since $ (\Delta_{V,h}+1)^{-1}$: $L^2(\mathbb T)\to L^2(\mathbb T)$ is compact, so is  $\mathsf B(\lambda)$. The Fredholm alternative then  implies that 
 $\mathsf T_h(\mu)$ is not invertible if and only if
 there exists $u\in L^2(\mathbb T)$, $u\neq 0$ such that $(1- \mathsf B(\lambda))u=0$, that is if and only if
  there exists $g\in H^2(\mathbb T)$, $g\neq 0$ such that $\mathsf T_h(\mu)g=0$. This concludes the proof of Lemma~\ref{lem:elemQ}. 
\ep
\medskip

\noindent 
According to Lemma~\ref{lem:elemQ},  for every $\mu\in\C$:
\begin{equation}\label{eq.TH}
\text{$\mu\in \sigma_d(\mathsf P_h)=\sigma(\mathsf P_h)$ if and only if   $\mu$ is  a singularity of $\lambda\mapsto \mathsf  T_h(\lambda)^{-1}$}
\end{equation}
and, for such a $\mu$, one has
\begin{equation}\label{eq.THbis}
\dim(\Ker (\mathsf P_h-\mu))=\dim(\Ker (\mathsf T_h(\mu)).
\end{equation}
%or equivalently if $\mu\in {\rm Ker } (\mathsf T_h)$. 
 To prove Theorems~\ref{th:doublewell} and \ref{th:main}, we will thus investigate the singularities of $\lambda\mapsto \mathsf  T_h(\lambda)^{-1}$ near~$0$. 
%Let us mention that according to Lemma~\ref{lem:elemQ} and in view of Theorems~\ref{th:1} and \ref{th:main}, the strategy now consists in~\textcolor{blue}{...}.
%
%{\bf attention, ici il n'y a pas equivalence entre les deux systemes. Quand on remonte on trouve 
%$d_V g-\lambda f\in ker(d_V^*)$ qui peut etre non trivial sur une vari\'et\'e. Mais comme on sait que $\mathsf P_h$ a le bon nombre de vp, ca gagnera quand meme.}
\subsection{Finite dimensional reduction, a Grushin problem} In this section, we show that the singularities of $\mathsf  T_h(\lambda)$ near $0$ are those of a  matrix valued holomorphic function $E_{-+}(\lambda)$. To this end, we will construct a 
so-called Grushin problem.  
\medskip

\noindent
To build the Grushin problem, we first need to recall known results on the low-lying spectrum of $\Delta_{V,h}$. 
From the early works of Witten \cite{Wi82} and Helffer-Sj\"ostrand \cite{HeSj85_01}, we know that there is a 
one-to-one
correspondence between the local minima of $V$ and the smallest eigenvalues of $\Delta_{V,h}$. This correspondence was further investigated by several authors and sharp asymptotic equivalents  of these small eigenvalues were finally obtained in \cite{BoGaKl05_01} and \cite{HeKlNi04_01} under assumptions on the relative positions of the minima and  saddle points, and in \cite{Mi19} in the general case. The following version gives the general form of these asymptotic equivalents  in a non-degenerate setting.

\begin{theorem}\label{th:vpWitten}\cite{BoGaKl05_01}, \cite{HeKlNi04_01}.  Let $U=-V$ be a Morse function.  There exist $\epsilon_*>0$, $C>0$,  and $h_0>0$ such that, for all $h\in]0,h_0]$, 
the nonnegative self-adjoint operator $(\Delta_{V,h},H^2(\mathbb T))$ admits exactly $n_0$ eigenvalues
(counted with algebraic multiplicity) in $[0,\epsilon_*h]$:
\begin{equation}\label{EKW1}
\sigma (\Delta_{V,h} ) \cap [0, \varepsilon_{*} h ]= \{0, \mu_{2,h}^\Delta, \ldots, \mu_{n_0,h}^\Delta \}, 
\end{equation}
where $0$ is a simple eigenvalue of $\Delta_{V,h}$. 
Let us order $\{\mu_{2,h}^\Delta, \ldots, \mu_{n_0,h}^\Delta\}$ such that $\mu_{j,h}^\Delta\le \mu_{j+1,h}^\Delta$ for $j=1,\ldots,n_0-1$. Then,  if~\eqref{eq.H2} holds, it holds for all  $j=2,\ldots,n_0$:
\begin{equation} \label{EKW2}
\mu_{j,h}^\Delta  =   a_h ( \m_j)\,h\, e^{- \frac 2h S ( \m_j ) }  ,
\end{equation}
where 
$S : \mathsf U^{(0)} \rightarrow ( 0 , + \infty ]$ is defined in \eqref{a27} (see also~\eqref{eq.order}) and $ a_h(\m)$  admits a classical expansion
 $ a_h(\m_j)\sim \sum_{k\geq 0}h^k a_k(\m_j)$ with 
\begin{equation} \label{EKW3}
 a_0 ( \m_j ) = \frac{1}{2 \pi}\sum_{\s \in {\bf j} ( \m_j )} \sqrt{|V'' ( \m_j )V'' ( \s )|}  ,
\end{equation}
  where ${\bf j} : \mathsf U^{( 0 )} \to \mathcal P ( \underline{\mathsf U}^{( 1 )} \cup \{ \s_{1} \} )$ is defined in \eqref{a26}.

\end{theorem}

\noindent
We are now in position to construct a Grushin problem. 
Set 
$$\Pi= \pi_{[0, \varepsilon_{*} h ]} (\Delta_{V,h} ) \text{ and } \mathsf E_0= {\rm Ran }(\Pi ),$$
where  $\pi_{[a,b]}(\Delta_{V,h} )$ is the spectral   projector associated with $\Delta_{V,h}$ and the interval $[a,b]$. 
According to Theorem~\ref{th:vpWitten},
 the space $\mathsf E_0$ has dimension $n_0$ for all $h>0$ small enough.  
 Let $(\Psi_j)_{j=1,\ldots n_0}$ be an  orthonormal basis of the space $\mathsf E_0$. %\textcolor{blue}{With a slice abuse of notation, we still denote $\Psi_j\mathsf 1_{\pm 1}$ and $\Psi_j$, for all $j$}.  
  We assume without loss of generality that $\Psi_1$ is proportional to $e^{-V/h}$. Introduce the operators
\be\label{eq:defR-}
\begin{split}
R_-:\;&\C^{n_0}\rightarrow L^2({\mathbb T})\\
&u\mapsto \sum_{j=1}^{n_0}u_j \Psi_j
\end{split}
\ee
and
\be\label{eq:defR+}
\begin{split}
R_+:\;&L^2({\mathbb T})\rightarrow\C^{n_0} \\
&u\mapsto (\<u ,\Psi_j\>_{L^2})_j.
\end{split}
\ee
We equip $\C^{n_0}$ with the $\ell_2$ norm. 
Notice  that $R_+R_-=\mathsf I_{\C^{n_0}}$ and $R_-R_+=\Pi$. In addition, $||R_+||\le 1$ and $||R_-||\le 1$, for all $h>0$. These inequalities will be used many times in what follows. 
From now on, we denote 
$$\widehat \Pi=\mathsf I-\Pi.$$
\begin{lemma}\label{eq:lemQhat}  Let $U$ be a Morse function. 
For any $\lambda\in\C$, the operator $\widehat {\mathsf T}_{h}(\lambda):=\widehat \Pi \mathsf  T_h(\lambda) \widehat \Pi$ acting on 
$\widehat\Pi L^2:=\widehat\Pi(L^2({\mathbb T}))$ with domain $D(\widehat {\mathsf T}_{h}(\lambda))=\{u\in \widehat\Pi L^2,\;u\in H^2({\mathbb T})\}$ is closed. Moreover, there exist $C,\epsilon_0, h_0>0$ such that
for all $h\in]0,h_0]$, and for all $\lambda\in D(0, \epsilon_0 h)$, $\widehat {\mathsf T}_{h}(\lambda)$ is invertible, holomorphic with respect to $\lambda$, and 
$
\Vert \widehat {\mathsf T}_{h}(\lambda)^{-1}\Vert\leq Ch^{-1}.
$
\end{lemma}
\noindent
\bp The proof is very close to the proof of Lemma 2.1 in \cite{LePMi20}. We sketch it for reader's convenience. 
We first observe that since $W$ is bounded, the operator $\mathsf  T_h(\lambda)$ with domain 
$D(\mathsf  T_h(\lambda))=D(\Delta_{V,h})=H^2({\mathbb T})$ is closed and densely defined. Its adjoint 
$\mathsf  T_h(\lambda)^*$ satisfies $D(\mathsf  T_h(\lambda)^*)=H^2({\mathbb T})$.
Suppose that 
$(u_n,\widehat {\mathsf T}_{h}(\lambda) u_n)\in D(\widehat {\mathsf T}_{h}(\lambda))\times L^2({\mathbb T})$ converges to $(u,v)\in L^2({\mathbb T})\times L^2({\mathbb T})$. For any 
$j=1,\ldots, n_0$, one has $\Psi_j \in  D(\Delta_{V,h})=D(\mathsf  T_h(\lambda)^*)$ and hence
for all $n\in\N$, one has
\be\label{eq:Qhat1}
\Pi \mathsf  T_h(\lambda) u_n=\sum_{j=1}^{n_0}\<\mathsf  T_h(\lambda) u_n,\Psi_j\> \Psi_j=\sum_{j=1}^{n_0}\< u_n,\mathsf  T_h(\lambda)^* \Psi_j\> \Psi_j.
\ee
Consequently, the sequence ($\Pi \mathsf  T_h(\lambda) u_n)_n$ converges and using  the identity
$$
\mathsf  T_h(\lambda) u_n=\Pi \mathsf  T_h(\lambda) u_n+\widehat {\mathsf T}_{h}(\lambda)  u_n,
$$
it follows that $(\mathsf  T_h(\lambda) u_n)$ converges. Since $\mathsf  T_h(\lambda)$ is closed as a bounded perturbation of a closed operator, and $u_n\in H^2(\mathbb T)=D(\mathsf  T_h(\lambda))$, one deduces that $u\in D(\mathsf  T_h(\lambda))$ and $\mathsf  T_h(\lambda) u=\lim \mathsf  T_h(\lambda) u_n$. Since $\widehat\Pi=1-\Pi$ is bounded, this implies that $u\in D(\widehat {\mathsf T}_{h}(\lambda))$ and 
$\widehat {\mathsf T}_{h}(\lambda) u=v$ which proves that $\widehat {\mathsf T}_{h}(\lambda)$ is closed. In addition, the operator $\widehat {\mathsf T}_{h}(\lambda)$ is clearly densely defined on $\widehat\Pi L^2$.
Let us now study the invertibility of $\widehat {\mathsf T}_{h}(\lambda)$. For any $u\in D(\widehat {\mathsf T}_{h}(\lambda))$, one has by definition
\begin{equation*}
\begin{split}
\Re\<\widehat {\mathsf T}_{h}(\lambda) u,u\>=\Re\< (\Delta_{V,h}-\lambda W+\lambda^2) \widehat \Pi u,\widehat \Pi u\>.
\end{split}
\end{equation*}
Using Theorem \ref{th:vpWitten}, this implies that for $|\lambda| <\epsilon_0h$, one has 
\begin{equation*}
\begin{split}
\Re\<\widehat {\mathsf T}_{h}(\lambda) u,u\>&\geq \epsilon_*h\Vert  \widehat \Pi u\Vert^2-\Re(\lambda)\<W  \widehat \Pi u, \widehat \Pi u\>+\Re(\lambda^2)\Vert  \widehat \Pi u\Vert^2\\
&\geq (\epsilon_*-\epsilon_0\Vert W\Vert_{L^\infty})h\Vert  \widehat \Pi u\Vert^2-\epsilon_0^2h^2\Vert  \widehat \Pi u\Vert^2\\
&\geq \frac {\epsilon_*} 2 h\Vert  \widehat \Pi u\Vert^2
\end{split}
\end{equation*}
for $\epsilon_0$ small enough. This proves that $\widehat {\mathsf T}_{h}(\lambda)$ is injective when $|\lambda| <\epsilon_0h$. We observe that the same proof shows that 
 $\widehat {\mathsf T}_{h}(\lambda)^*=\widehat \Pi \mathsf  T_h(\lambda)^*\widehat \Pi$ is injective. 
 Moreover, Cauchy-Schwarz inequality implies that for all $u\in D(\widehat {\mathsf T}_{h}(\lambda))$, 
\be\label{eq:Qhat2}
 \Vert \widehat {\mathsf T}_{h}(\lambda) u\Vert \geq \frac {\epsilon_*} 2 h\Vert  \widehat \Pi u\Vert.
 \ee
 Let us now prove that $\widehat {\mathsf T}_{h}(\lambda)$ is surjective if $|\lambda| <\epsilon_0h$. We first observe that $\Ran(\widehat {\mathsf  T}_h(\lambda))$ is closed
 since $\widehat {\mathsf T}_{h}(\lambda)$ is closed and   the convergence of any sequence $(\widehat {\mathsf T}_{h}(\lambda) u_n)$  implies the convergence of $(u_n)$  thanks to \eqref{eq:Qhat2}.
 Hence it is sufficient to prove that $\Ran (\widehat {\mathsf T}_{h}(\lambda))^\bot =\{0\}$. Suppose that
 $v\in L^2(\mathbb T)$ satisfies
 $\<\widehat {\mathsf T}_{h}(\lambda) u,v\>=0$
 for all $u\in D(\widehat {\mathsf T}_{h}(\lambda))$. Then 
 $$
 \<\mathsf  T_h(\lambda) \widehat \Pi u,\widehat \Pi v\>=0
 $$
 which implies that $\widehat\Pi v\in D(\mathsf  T_h(\lambda)^*)$ and 
 $
 \<u,\widehat \Pi \mathsf  T_h(\lambda)^*\widehat\Pi v\>=0
 $
 for all $u\in D(\widehat {\mathsf T}_{h}(\lambda))$ which is dense in $\widehat\Pi L^2$. Hence  $\widehat \Pi \mathsf  T_h(\lambda)^*\widehat\Pi v=0$ and since $\widehat {\mathsf T}_{h}(\lambda)^*$ is injective, this implies that $v=0$. This ends the proof of the lemma. 
\ep

%\textcolor{blue}{ 
%Dorian : il me semble que l'on ne l'utilise plus avec la version Rouch\'e-Michel :
%\begin{remark}\label{re.t-1}
%Notice that by Lemma~\ref{eq:lemQhat}, and since   $\partial_\lambda\widehat {\mathsf T}_{h}(\lambda)^{-1}=-  \widehat {\mathsf T}_{h}(\lambda)^{-1}   \partial_\lambda\widehat {\mathsf T}_{h}(\lambda) \widehat {\mathsf T}_{h}(\lambda)^{-1}$, it holds for all $h$ small enough and all  $\lambda\in D(0, \epsilon_0 h)$,  
%$
%\Vert \partial_\lambda\widehat {\mathsf T}_{h}(\lambda)^{-1}\Vert=O(h^{-2}).
%$ 
%\end{remark}}
\noindent
We now  introduce  the Grushin operator
\be\label{eq:defGrushin}
\mathcal K_h(\lambda)=
\left(\begin{array}{cc}
\mathsf  T_h(\lambda)&R_-\\
R_+&0
\end{array}
\right).
\ee

\begin{proposition}\label{prop:grushin}  Let $U$ be a Morse function.  There exist $\epsilon_0>0$ and $h_0>0$ such that, for all $h\in]0,h_0]$ and all $\lambda\in D(0,\epsilon_0 h)$, the operator
$\mathcal K_h(\lambda)$ is invertible. Moreover, its inverse $\eee_h(\lambda)$ writes
$$
\eee_h(\lambda)=
\left(\begin{array}{cc}
E(\lambda)&E_+(\lambda)\\
E_-(\lambda)&E_{-+}(\lambda)
\end{array}
\right),
$$
where $E,E_-,E_+,E_{-+}$ are holomorphic in $D(0, \epsilon_0h )$
and satisfy the following  formulas:
$$
E_+(\lambda)\ =\ R_--\widehat {\mathsf T}_{h}(\lambda)^{-1}\widehat\Pi \mathsf  T_h(\lambda) R_-\,,\ \ E_-(\lambda)\ =\ R_+-R_+\mathsf  T_h(\lambda) \widehat {\mathsf T}_{h}(\lambda)^{-1}\widehat\Pi \,,
$$
\be\label{eq:grushin2}
E_{-+}(\lambda)\ =\ -R_+\mathsf  T_h(\lambda) R_-+R_+\mathsf  T_h(\lambda) \widehat\Pi \widehat {\mathsf T}_{h}(\lambda)^{-1}\widehat\Pi \mathsf  T_h(\lambda) R_-: \C^{n_0}\to \C^{n_0},
\ee
and 
\be\label{eq:grushin3}
E(\lambda)\ =\ \widehat {\mathsf T}_{h}(\lambda)^{-1}\widehat\Pi\,.
\ee
Moreover, for $\lambda\in D(0,\epsilon_0 h)$, $\mathsf  T_h(\lambda)$ is invertible if and only if the matrix $E_{-+}(\lambda)$ is invertible, in which case
it holds
$$
\mathsf  T_h(\lambda)^{-1}\ =\ E(\lambda)-E_+(\lambda)E_{-+}^{-1}(\lambda)E_-(\lambda).
$$
\end{proposition}
\noindent
\bp
Thanks to Lemma \ref{eq:lemQhat}, the proof is reduced to an algebraic computation which is completely analogous to
the one used in the  proof of  \cite[Lemma 2.2]{LePMi20}.
\ep
\medskip

\noindent
We now give some direct consequences of Proposition~\ref{prop:grushin} and Lemma~\ref{eq:lemQhat}, which will be used in the following.  
By definition, one has $R_+ \mathsf  T_h(\lambda) \widehat \Pi=-\lambda R_+ W\widehat\Pi$. Hence \eqref{eq:grushin2} becomes
$$
E_{-+}(\lambda)\ =\ -R_+\mathsf  T_h(\lambda) R_-+\lambda^2R_+W \widehat\Pi \widehat {\mathsf T}_{h}(\lambda)^{-1}\widehat\Pi W R_-.
$$
Introducing the matrices 
\be\label{eq:defMV}
\mathsf M_{V,h}=R_+ \Delta_{V,h} R_-, \ \mathsf W_h=R_+W R_-, \text{ and } \mathsf G_h(\lambda)=\lambda^2(\mathsf I_{\C^{n_0}}-R_+W \widehat\Pi \widehat {\mathsf T}_{h}(\lambda)^{-1}\widehat\Pi W R_-),
\ee
this rewrites
\be\label{eq:matEmp}
-E_{-+}(\lambda)\ =\mathsf M_{V,h}-\lambda\mathsf W_h+\mathsf G_h(\lambda).
\ee
Moreover, it follows from Lemma \ref{eq:lemQhat} that $\lambda\in D(0, \epsilon_0 h)\mapsto \mathsf G_h(\lambda)$ is holomorphic   and that there exists $C>0$ such that for all $h$ small enough and all $\lambda\in D(0, \epsilon_0 h)$,
\be\label{eq:matEmp2} 
 |\mathsf G_h(\lambda)|\le C|\lambda|^2h^{-1}.
 \ee 
  According to \eqref{eq.TH} and to Proposition~\ref{prop:grushin}, 
\begin{equation}\label{eq.start}
\text{$\lambda \in D(0, \epsilon_0 h)\cap \sigma(\mathsf P_h)$ \ \ if and only if\ \  $\lambda\in D(0, \epsilon_0 h)$ is a singularity of $E_{-+}^{-1}(\lambda)$.} 
\end{equation}
To prove Theorems \ref{th:doublewell} and \ref{th:main}, the strategy  consists in  studying the singularities of $E_{-+}^{-1}(\lambda)$ in  $D(0, \epsilon_0 h)$, which first requires to compute asymptotic equivalents of $\mathsf M_{V,h}$ and $\mathsf W_h$ as $h\to 0$ (see~\eqref{eq:matEmp}}). 
%The one for $\mathsf M_{V,h}$ is not difficult to obtain, by definition of $R_+$ and $R_-$. 
%Then,  by definition of $\mathsf M_{V,h}$, it holds for $h$ small enough:
%\begin{equation}\label{eq.MH}
%\mathsf M_{V,h}={\rm diag}(0, \mu_{2,h}^\Delta,\ldots, \mu_{h}^\Delta(\m_{n_0})).
%\end{equation}
%\textcolor{blue}{pas vrai}
%Notice that so far we have never used~\eqref{eq.H1}. 
 %For ease of notation, we set for all $j\in \{1,\ldots,n\}$
 %$$S_j= S ( \m_{j} ) \text{ and } \mu_j(h)= \mu(\m_j,h).$$
%In all what follows we keep this labeling. 

\section{The double well case}
\noindent
In this section we prove Theorem \ref{th:doublewell}. To this end, we assume throughout this section  that    $\mathsf U^{(0)}$ has exactly two elements $\m_1$ and $\m_2$ such that $V(\m_1)<V(\m_2)$. Assume also that the two elements $\s_1,\s_2$ of 
$\mathsf U^{(1)}$ satisfy $V(\s_1)>V(\s_2)$.  Recall that in this case, one has  by construction of $S$   (see~\eqref{a27})  and $\bf j$ (see~\eqref{a26}), $S(\m_2)=V(\s_2)-V(\m_2)$  and $\bf j(\m_2)=\{s_2\}$ (see Remark~\ref{re.REDW}). Thus, by Theorem~\ref{th:vpWitten}, it holds
\begin{equation}\label{eq.WDW}
\mu_{2,h}^\Delta=a_h ( \m_2)\,h\, e^{- \frac{2 (V(\s_2)-V(\m_2))}{h}},
\end{equation}
with  $ a_h(\m_2)  \sim \sum_{k\geq 0}h^k a_k(\m_2)$ and 
$ a_0 ( \m_2 ) = \frac{1}{2 \pi} \sqrt{|V'' ( \m_2 )V'' ( \s_2 )|}$. 
 \subsection{Asymptotic equivalent of   $\mathsf W_h$, as $h\to 0$}  
%Thanks to Lemma \ref{eq:defQlambda}, we know that any eigenvalue $\lambda(h)$ of $\mathsf P_h$ is a singularity of 
%$\mathsf  T_h(\lambda)$. Recall that by Theorem \ref{th:1},   
%$\mathsf P_h$ has exactly two eigenvalues $\lambda_{1,h},\lambda_{2,h}$ in $\{\Re(z)<\epsilon_0 h^2\}$. Consequently, in order to prove
%Theorem \ref{th:doublewell}, it suffices to show  that there exists $\epsilon>0$ such that $\mathsf  T_h(\lambda)$ has exactly two singularities $\lambda_{1,h},\lambda_{2,h}$ in $D(0,\epsilon h)$ and 
%that 
%\be\label{eq:proof2D1}
%\lambda_{1,h}=0\text{ and }\lambda_{2,h}=\zeta(\m_2,h)e^{-S_2/h} \text{ as $h\to 0$}, 
%\ee
%where  $\zeta(\m_2,h)$ and $S_2$ are defined  in Theorem  \ref{th:doublewell}. Thanks to Proposition \ref{prop:grushin}, $\lambda\in D(0,\epsilon_0 h)\mapsto  \mathsf  T_h(\lambda)^{-1}$ and $\lambda\in D(0,\epsilon_0 h)\mapsto E_{-+}^{-1}(\lambda)$ have the same poles. 
%On the other hand,  since $n_0=2$ and $V(\s_1)>V(\s_2)$, by Theorem \ref{th:vpWitten}, $\Delta_{V,h}$ admits two eigenvalues $\mu_1(h),\mu_2(h)$ in $[0,\epsilon_*h]$ with 
%$\mu_1(h)=0$ and 
%\be\label{eq:asympmu2}
%\mu_2(\m_2,h)=a (\m_2,h) e^{- \frac{2 S_2}{h}} 
%\ee
%where $S_2=V(\s_2)-V(\m_2)$ and 
%$ a_2\sim h\sum_{k\geq 0}h^k a_{2,k}$ with 
%\begin{equation} \label{e13}
% a_0(\m_2) = \frac{1}{2 \pi} \sqrt{|V'' ( \m_2 )V'' ( \s_2 )|} .
%\end{equation}
% 
%
%\noindent
To compute an asymptotic equivalent of   $\mathsf W_h$ in the limit $h\to 0$, we first need to define   $(\Psi_1,\Psi_2)$ with the help of so-called quasi-modes, where we recall that  $(\Psi_1,\Psi_2)$ is   an orthonormal basis of eigenvectors of $\Delta_{V,h}$ associated respectively with the eigenvalues $0$ and $\mu_{h}(\m_2)$. First of all, we choose for $h>0$,
 $$  \Psi_1=\frac 1{Z_{1,h}}e^{-(V-V(\m_1))/h} \text{ with } Z_{1,h} =\Vert e^{-(V-V(\m_1))/h}\Vert_{L^2(\mathbb T)},$$
 where by Laplace's method:   $Z_{1,h} \in \eee_{cl} ((\pi h /V''(\m_1))^{1/4})$ (since $\m_1$ is the unique global minimum of $V$ on $\mathbb T$).
  Because $\Delta_{V,h}\Psi_1=0$, $\Psi_1\in \mathsf E_0$. We then construct $\Psi_2$ as follows. Define 
\be\label{eq:deff2}
\varphi_2= \frac 1{Z_{2,h}}\chi_2e^{-(V-V(\m_2))/h} \text{ with } Z_{2,h}=\Vert e^{-(V-V(\m_2))/h}\Vert_{L^2(\mathbb T)}, 
\ee
and where    $\chi_2\in \mathcal C_c^\infty({\mathbb T},[0,1])$   satisfies  $\indic_{B(\m_2,r)}\leq \chi_2\leq \indic_{B(\m_2,2r)}$,  and where   $r>0$. If $r>0$ is small enough, $\m_2$ is the unique global minimum of $V$ on the closure of $B(\m_2,2r)$ (because $V$ is a Morse function). Thus, by Laplace's method: $Z_{2,h} \in\eee_{cl} ((\pi h /V''(\m_2))^{1/4})$. 
In addition, for such fixed $r>0$,    
$V>V(\m_2)$ on $\supp(\nabla\chi_2)$, and therefore, there exists $C>0$ such that for $h$ small enough:
\be\label{eq:estimDf_2}
\Delta_{V,h} \varphi_2=O(e^{-C/h}) \text{ in $L^2({\mathbb T})$.} 
\ee
 On the other hand, since $\Delta_{V,h}$ is self-adjoint, it follows from the localisation of the spectrum in Theorem \ref{th:vpWitten} (with $n_0=2$ there) that 
\be\label{eq:estimresW}
\forall z\in\partial D(0, {\epsilon_*}  h/2),\;(\Delta_{V,h}-z)^{-1}=O(h^{-1}).
\ee
By definition of $\Pi$ and by Theorem~\ref{th:vpWitten},  
$$
\Pi=\frac 1{2i\pi}\int_{\partial D(0,\frac{\epsilon_*} 2 h)}(z-\Delta_{V,h})^{-1}dz
$$
and using \eqref{eq:estimDf_2} and \eqref{eq:estimresW}, it follows that  $\Pi\varphi_2$ satisfies
\be\label{eq:approxe21}
\Pi\varphi_2-\varphi_2=\frac {1}{2i\pi}\int_{\partial D(0,\frac{\epsilon_*} 2 h)}z^{-1}(z-\Delta_{V,h})^{-1}\Delta_{V,h} \varphi_2dz=O(e^{-C/h}).
\ee
Moreover, since $V(\m_2)>V(\m_1)$ and $\Pi\Psi_1=\Psi_1$, one has for $h$ small enough:
\be\label{eq:approxe22}
 \<\Pi\varphi_2,\Psi_1\>_{L^2(\mathbb T)}=\<\varphi_2,\Psi_1\>_{L^2(\mathbb T)}=O(e^{-C/h}).
\ee
We finally set 
$$
  \Psi_2=\frac{\Pi\varphi_2-\<\Pi\varphi_2,\Psi_1\>_{L^2(\mathbb T)}\Psi_1}{\Vert \Pi\varphi_2-\<\Pi\varphi_2,\Psi_1\>\Psi_1\Vert_{L^2(\mathbb T)}}.
$$
The function $  \Psi_2$ belongs to $\mathsf E_{0}$,  is orthogonal to $\Psi_1$, and  $\Vert \Psi_2\Vert_{L^2(\mathbb T)}=1$. From now on, we consider $(\Psi_1,\Psi_2)$ constructed as above, as a orthonormal basis of $\mathsf E_0$.   
Notice that, using \eqref{eq:approxe21} and \eqref{eq:approxe22}, one has 
\be\label{eq:approxe23}
\Psi_2=\varphi_2+O(e^{-C/h}) \text{ in $L^2({\mathbb T})$.}
\ee

\begin{lemma}\label{lem:compW-2D}   
With the above choice of $\Psi_1,\Psi_2$, there exists $C>0$ such that  for all $h$ small enough,
$$
\mathsf W_h=
\left(
\begin{array}{cc}
\gamma_{1,h}&0\\
0&\gamma_{2,h}
\end{array}
\right)
+O(e^{-C/h}),
$$
 where  we recall that $\mathsf W_h$ is defined in~\eqref{eq:defMV} and where, for $i\in\{1,2\}$, $\gamma_{i,h}$ satisfies as $h\to 0$: $
\gamma_{i,h}\sim  \sum_{k\geq 0}h^k\gamma_{\alpha, k}(\m_i)
+\sqrt h \sum_{k\geq 0}h^k\gamma_{V,k}(\m_i)$,
  with $$\text{$\gamma_{\alpha, 0}(\m_i)=\alpha(\m_i)$ and 
$\gamma_{V,0}(\m_i)=2\sqrt{\frac{V''(\m_i)}\pi}$.}$$
\end{lemma}
\noindent
\bp
Since $(\Psi_1,\Psi_2)$ is an orthonormal family, one has $\mathsf W_h=(\<W \Psi_i,\Psi_j\>_{L^2(\mathbb T)})_{i,j=1,2}$. 
In addition,  $\Psi_1=O(e^{-C/h})$ in $L^2(\supp(\varphi_2))$. Hence, using~\eqref{eq:approxe23},  for all $i\neq j$, $\<W \Psi_i,\Psi_j\>_{L^2(\mathbb T)}=O(e^{-C/h})$.
Suppose now that $i\in\{1,2\}$ is fixed. By definition of $W$, one has
$\<W \Psi_i,\Psi_i\>_{L^2(\mathbb T)}=\<\alpha \Psi_i, \Psi_i\>_{L^2(\mathbb T)}+2\<|\partial_xV| \Psi_i,\Psi_i\>_{L^2(\mathbb T)}$.
By definition of $\Psi_1$ and $\Psi_2$ above, using    
Laplace's method, one has as $h\to 0$:   $\<\alpha \Psi_i,\Psi_i\>_{L^2(\mathbb T)}\sim \sum_{k\geq 0}h^k\gamma_{\alpha, k}(\m_i)$ with $\gamma_{\alpha,0}(\m_{i})=\alpha(\m_{i})$.
On the other hand, for $i\in\{1,2\}$, one has for $\delta>0$ small enough,
\begin{equation*}
\begin{split}
\<|\partial_xV| \Psi_i,\Psi_i\>_{L^2(\mathbb T)}&=Z_{i,h}^{-2}\int_{|x-\m_i|<\delta} |\partial_xV(x)| e^{-2(V(x)-V(\m_i))/h}dx+O(e^{-c/h})\\
&=Z_{i,h}^{-2}\Big(\int_{\m_i}^{\m_i+\delta}\partial_xV(x) e^{-2(V(x)-V(\m_i))/h}dx\\
&\phantom{****}-\int_{\m_i-\delta}^{\m_i} \partial_xV(x) e^{-2(V(x)-V(\m_i))/h}dx
\Big)
+O(e^{-c/h})\\
&=hZ_{i,h}^{-2}+O(e^{-c/h}).
\end{split}
\end{equation*}
Using the fact that $Z_{i,h} \in\eee_{cl} ((\pi h /V''(\m_i))^{1/4})$, this ends the proof of the lemma. 
\ep

\noindent
We are now in position to prove Theorem \ref{th:doublewell}. 
\subsection{End of the proof of Theorem \ref{th:doublewell}}
  Recall that the strategy consists in localizing  the singularities of $E_{-+}^{-1}(\lambda)$ in $D(0, \epsilon_0 h)$ (see \eqref{eq.start}). First of all,    $\lambda=0$ is always a singularity of $\lambda\in  D(0, \epsilon_0 h)  \mapsto E_{-+}^{-1}(\lambda)$, since $\mathsf T_h(0)\Psi_1=0$ (see~Proposition~\ref{prop:grushin}). Let us now  look for the other singularities of  $\lambda\in  D(0, \epsilon_0 h)  \mapsto E_{-+}^{-1}(\lambda)$.  
 Since $\dim\mathsf E_0=2$,  $\Delta_{V,h}\Psi_1=0$, and $\Psi_2\in \mathsf E_0$ is orthogonal to $\Psi_1$, it holds: $\Delta_{V,h}\Psi_2=\mu_{2,h}^\Delta\Psi_2$ and therefore, $\mathsf M_{V,h}=\diag( 0,  \mu_{2,h}^\Delta)$. 
By Lemma~\ref{lem:compW-2D}, \eqref{eq:matEmp}, and \eqref{eq:matEmp2},  one then has
$$
-E_{-+}(\lambda)=
\left(
\begin{array}{cc}
0&0\\
0&\mu_{2,h}^\Delta
\end{array}
\right)
-\lambda
\left(
\begin{array}{cc}
\gamma_{1,h}&0\\
0&\gamma_{2,h}
\end{array}
\right)
+\mathsf r_h(\lambda),
$$
where $ \lambda\in D(0, \epsilon_0 h)\mapsto \mathsf r_h(\lambda)$ is holomorphic and $  \mathsf r_h(\lambda)=O(\lambda^2h^{-1}+\lambda e^{-C/h})$ for all $\lambda\in D(0, \epsilon_0 h)$ and $h$ small enough.  
Set   $$\Gamma_h=\diag(\gamma_{1,h},\gamma_{2,h}).$$ 
%Assume that~\eqref{eq.H1} holds. 
Then, by Lemma~\ref{lem:compW-2D}:
%and since $\alpha(\m_i)\neq 0$ (see indeed~\eqref{eq.H1-c}), it holds
\begin{equation}\label{eq.Gam}
\Gamma_h^{-1}=O( h^{-1/2} ).
\end{equation}
%\textcolor{blue}{$O(1)$ si $\alpha(\m_i)\neq 0$ (ce qui est le cas ici)}. 
Therefore, one deduces that
 \begin{equation}\label{eq.EE+--}
-E_{-+}(\lambda)= \Gamma_h \big(F_h(\lambda)
+\mathsf R_h(\lambda)\big) \text{ with } F_h(\lambda)=\left(
\begin{array}{cc}
-\lambda&0\\
0&\mu_{2,h}^\Delta/\gamma_{2,h}-\lambda
\end{array}
\right),
\end{equation}
where $\mathsf R_h(\lambda)$ is holomorphic with respect to $\lambda\in  D(0, \epsilon_0 h)$ and $\mathsf R_h(\lambda)=O(\lambda^2h^{-\frac 32}+\lambda h^{-\frac 12} e^{-C/h})$  for all $\lambda\in D(0, \epsilon_0 h)$ and $h$ small enough. 
Set  $$\eta_{2,h}:=\mu_{2,h}^\Delta/\gamma_{2,h}.$$
According to  Lemma \ref{lem:compW-2D}, to~\eqref{eq.WDW}, and to the relation $\alpha(\m_2)>0$ implied by \eqref{eq.H1} (see indeed \eqref{eq.H1-c}), one has in the limit $h\to 0$:
\begin{equation}
\label{eq.DLeta}
\eta_{2,h}= \zeta_h(\m_2)\,h\,e^{-\frac 2h (V(\s_2)-V(\m_2))},
\end{equation}
where  $\zeta_h(\m_2)\sim \sum_{k\geq 0}h^{\frac k 2}\zeta_k(\m_2)$ and   $
\zeta_0(\m_2)=\frac 1{2\pi} \sqrt{|V''(\m_2)V''(\s_2)|}/\alpha(\m_2)$.
Let $K\ge 2$ be fixed in  what follows. 
Set 
$$\ddd_{K}=\big\{\lambda\in\C,\;|\lambda-\eta_{2,h}|<h^Ke^{-\frac 2h (V(\s_2)-V(\m_2))}\big\},$$  whose closure  is
by~\eqref{eq.DLeta}
 included in $D(0, \epsilon_0 h)$ for $h$ small enough. 
%Recall that by~\eqref{eq.WDW}, one has $\mu_{2,h}^\Delta = ha_0(\m_2)e^{-\frac 2h(V(\s_2)-V(\m_2))}(1+o(1))$ as $h\to 0$.
 Hence, for any $\lambda\in \overline{\ddd_{K}}$,  since $K\ge 2$, one has, for all $h$ small enough,  
 \begin{equation}
\label{eq.lambda2}
C^{-1}he^{-\frac 2h(V(\s_2)-V(\m_2))} \le \vert \lambda\vert \le Che^{-\frac 2h(V(\s_2)-V(\m_2))}.
 \end{equation}
  Consequently, for any $\lambda\in \partial\ddd_{K}$, 
 the matrix
$F_h(\lambda)$ is invertible and 
\begin{equation}\label{eq.F-1}
F_h(\lambda)^{-1}=O(h^{-K}e^{\frac 2h(V(\s_2)-V(\m_2)) }) \text{ \ on $ \partial\ddd_{K}$}.
\end{equation}
 On the other hand,  it follows from the estimate on $\mathsf R_h$ below \eqref{eq.EE+--} that  for any $\lambda\in \partial\ddd_{K}$, 
 %one has
 %since 
 %$|\lambda|\leq Ch e^{-(2(V(\s_2)-V(\m_2)))/h}$ (for some fixed $C>0$), one has 
$$\mathsf R_h(\lambda)=O( \sqrt h \, e^{-4(V(\s_2)-V(\m_2))/h}+  \sqrt h \,  e^{-(2(V(\s_2)-V(\m_2))+C)/h})=O(e^{-(2(V(\s_2)-V(\m_2))+C)/h}).$$
 Hence, by~\eqref{eq.EE+--},
$E_{-+}(\lambda)$ is invertible  on $ \partial\ddd_{K}$ for all $h$ small enough  and 
$$
-E_{-+}^{-1}(\lambda)= \big(1+O(e^{-\frac C{2h}})\big)F_h(\lambda)^{-1}\Gamma_h^{-1}
%= \big(F_h(\lambda)^{-1}+  O(e^{-\frac C{4h}})\big)\Gamma_h^{-1} \,  
\text{  on $ \partial\ddd_{K}$},
$$
where we have used~\eqref{eq.F-1} and the invertibility of $\Gamma_h$. 
Using in addition $\Vert F_h(\lambda)^{-1}\Vert\,|\partial\ddd_{K}|=O(1)$ and \eqref{eq.Gam}, this implies that for all $h$ small enough:
\begin{equation}
\label{eq.tour}
\begin{split}
\frac1{2i\pi}\int_{\partial\ddd_{K}}E_{-+}^{-1}(\lambda)d\lambda
&= -\Big[\frac1{2i\pi}\int_{\partial\ddd_{K}}
\left(
\begin{array}{cc}
-\lambda&0\\
0&  \eta_{2,h}-\lambda
\end{array}
\right)^{-1}
d\lambda \Big]\, \Gamma_h^{-1}
+\Vert \Gamma_h^{-1}\Vert O(e^{-\frac C{2h}})\\
%\nonumber
&=\left(\begin{array}{cc}
0&0\\
0&\gamma_{2,h}^{-1}
\end{array}
\right)
+O(e^{-\frac C{4h}}) \text{ is non trivial.}
\end{split}
\end{equation}
 Hence, for all $K\ge 2$,  $\lambda\mapsto E_{-+}^{-1}(\lambda)$ admits at least a singularity $\alpha_{h,K}$  in the disk $\ddd_{K}$. In particular, one has for all $h$ small enough,
$$\alpha_{h,K}=  \eta_{2,h}+O(h^Ke^{-\frac 2h (V(\s_2)-V(\m_2))}), \text{ where $  \eta_{2,h}$ satisfies \eqref{eq.DLeta}}.$$ 
But, since \eqref{eq.H1} holds, Theorem~\ref{th:1} implies that for all $h$ small enough, $\sigma(\mathsf  P_h)\cap \{ {\rm Re}(z)\le  \epsilon_0 h^2\}$ is composed of two elements $0$ and $\lambda_{2,h}$. 
According to \eqref{eq.start},
the nonzero eigenvalue $\lambda_{2,h}$ then necessarily satisfies
$\lambda_{2,h}=\alpha_{h,K}$ for all $K\ge 2$. This ends the proof of Theorem~\ref{th:doublewell}. 
%  Hence, for all $K\ge 1$,  $\lambda\mapsto E_{-+}^{-1}(\lambda)(\lambda)$ has at least a singularity $\alpha_h$  in the disk $\ddd_{K}$. In particular, one has for $h$ small enough
%$$\alpha_h=  \eta_{2,h}+O(h^Ke^{-\frac 2h (V(\s_2)-V(\m_2))}), \text{ where $  \eta_{2,h}$ satisfies \eqref{eq.DLeta}}.$$ 
% Furthermore, if~\eqref{eq.H1} holds, by Theorem~\ref{th:1}, $\sigma(\mathsf  P_h)\cap \{ {\rm Re}(z)\le  \epsilon_0 h^2\}$ is composed of two elements.   For $h$ small enough, these two eigenvalues are  $0$ and $\alpha_h$, by~\eqref{eq.start} and since $\alpha_h\in \{ {\rm Re}(z)\le  \epsilon_0 h^2\}$ for $h$ small enough. This ends the proof of Theorem~\ref{th:doublewell}. 

\section{The multiple well case}
\noindent
In this section, we prove Theorem~\ref{th:main}. We assume that $U$ is a Morse function and  that  \eqref{eq.H2} is satisfied. Recall that 
the local minima $\m_{1} , \ldots , \m_{n_{0}}$ of $V$ are labeled such that  $( S ( \m_{j} ) )_{j \in \{ 1 , \dots , n_{0} \}}$ is decreasing (see~\eqref{eq.order}). We then  label $\{\Psi_1,\ldots,\Psi_{n_0}\}$, the basis of $\mathsf E_0$,  accordingly.

\subsection{An adapted basis of quasimodes}
In order to compute the matrices $\mathsf M_{V,h}$ and $\mathsf W_h$ accurately,  we will  
build  the basis $(\Psi_j)_{j=1,\ldots,n_0}$ of $\mathsf E_0$
from a family of quasi-modes $(\varphi_j)_{j=1,\ldots,n_0}$ constructed in \cite{BoLePMi22}.
First,    as in the previous section, set,  for $h>0$,
 $$  \varphi_1=\frac1{Z_{1,h}}e^{-(V-V(\m_1))/h} \text{ with } Z_{1,h} =\Vert e^{-(V-V(\m_1))/h}\Vert_{L^2(\mathbb T)} \in 
 \eee_{cl} (( \pi h/V''(\m_1))^{1/4}),$$
 and $\Psi_1=\varphi_1$. 
For any $j=2,\ldots,n_0$, let $\varphi_j$ be the $L^2(\mathbb T)$-normalized quasi-mode associated with $\m_j$ given in  \cite[Definition 4.3]{BoLePMi22} applied to the case of the Witten Laplacian $\Delta_{V,h} $.  
Since $\varphi_i$ is by definition supported in a neighborhood of $\mathsf C(\m_{i})$  (see~\cite[Section 4]{BoLePMi22}),
it follows from the second item of Assumption \eqref{eq.H2} that,  for all $i\in\{1,\ldots, n_0\}$ and $r>0$ small enough, one has for all $h$ small enough:  
%\argmin_{\text{supp }\varphi_j}f=\{\m_j\} \text{ and }
\be\label{eq:asympvarphim}
  \varphi_i= \frac{\mathsf 1_{D(\m_i,r)}}{Z_{i,h} }e^{-(V-V(\m_i))/h}+O(e^{-c/h} )\, \text{ in } L^2(\mathbb T), 
  %\text{ and } 
%\Vert (1-\Pi)\varphi_i\Vert_{L^2(\mathbb T)}= O(e^{-c/h})  
%\  \text{ and } \argmin_{\text{supp }\varphi_j}f=\{\m_j\},
%\  \varphi_i=Z_{i,h}e^{-(V-V(\m_i))/h} 
\ee 
%and $\varphi_j=$ 
where $Z_{i,h}\in \eee_{cl} ((\pi h /V''(\m_i))^{1/4})$. 
 For $j \in \{ 1 , \ldots , n_{0} \}$, we set
$$
  \kappa_{j,h}: = \< \Delta_{V,h} \varphi_j, \varphi_j\> .
$$
We have obviously $  \kappa_{1,h}=0$ and,  from   \cite[Proposition 5.1]{BoLePMi22}, we have for all $j \in \{ 2 , \ldots , n_{0} \}$,
\begin{equation} \label{a70}
\kappa_{j,h} \in \eee_{cl} \Big(  h e^{- 2S ( \m_{j} )  / h} \sum_{\s \in {\bf j} ( \m_{j} )} \frac{ |V''(\m_j)V''(\s)|^{\frac 12}}{2 \pi} \Big).
\end{equation}
The basis $(\Psi_j)_{j=1,\ldots,n_0}$ of $\mathsf E_0=\text{Ran } \Pi$ is then constructed from the family $(\varphi_j)_{j=1,\ldots,n_0}$   by the following procedure.   Set, for $j=1,\ldots,n_0$:  $v_j=\Pi \varphi_j$. For $h$ small enough, the family  $(v_j)_{j=1,\ldots,n_0}$ is then a basis of $\mathsf E_0$ (since, according to  \cite[Proposition 5.3]{BoLePMi22}, $\langle v_j,v_i\rangle_{L^2(\mathbb T)}=\delta_{i,j}+O(e^{-c/h})$, for all $i,j=1,\ldots,n_0$). We then consider 
the family $(\Psi_j)_{j=1,\ldots,n_0}$ obtained from  $(v_j)_{j=1,\ldots,n_0}$ by a Gram-Schmidt procedure as in \cite[page~30]{BoLePMi22}. 
According to  \cite[Proposition 5.3 and Eq. (5.15)]{BoLePMi22},
this orthonormal basis of $\mathsf E_0$  satisfies the following properties:
\begin{equation} \label{a71}
\text{for all $j,k=1,\ldots,n_0$}, \ \ \ \< \Delta_{V,h} \Psi_{j} , \Psi_{k} \> = \delta_{j , k}  \kappa_{j,h} +O\big(h^\infty \sqrt{ \kappa_j(h) \kappa_k(h)}\big) 
\end{equation}
and 
\begin{equation} \label{a72}
\text{for all $j=1,\ldots,n_0$}, \ \ \ \Psi_j=\varphi_j+O(e^{-c/h}) \text{ in } L^2(\mathbb T).
\end{equation}
Working in the  basis $(\Psi_j)_{j=1,\ldots,n_0}$ of $\mathsf E_0$, we obtain the following asymptotic equivalent of $\mathsf W_h$ (see~\eqref{eq:defMV}).
\begin{lemma}\label{lem:compW}
There exists $C>0$ such that the  matrix $\mathsf W_h$ satisfies, for $h$ small enough,
$$
\mathsf W_h=\Gamma_h+O(e^{-C/h}),
$$
where  $\Gamma_h=\diag( \gamma_{1,h},\ldots,\gamma_{n_0,h} )$ and, for  $j=1,\ldots,n_0 $:
$$
\gamma_{j,h}\sim  \sum_{k\geq 0}h^k\gamma_{\alpha, k}(\m_j)
+\sqrt h \sum_{k\geq 0}h^k\gamma_{V,k}(\m_j)
$$
with $\gamma_{\alpha, 0}(\m_j)=\alpha(\m_j)$, and 
$\gamma_{V,0}(\m_j)=2\sqrt{\frac{V''(\m_j)}\pi}$.

\end{lemma}
\noindent
\bp The asymptotic computation of 
$\<W\Psi_i,\Psi_i\>$  as $h\to 0$ is exactly the same as in Lemma \ref{lem:compW-2D}, using \eqref{eq:asympvarphim} and \eqref{a72}. The only point to be checked is that for any 
$i\neq j$, one has $\<W\Psi_i,\Psi_j\>=O(e^{-C/h})$. Thanks to \eqref{a72}, this is equivalent to say that 
for all $i\neq j$, one has $\<W\varphi_i,\varphi_j\>=O(e^{-C/h})$, which follows directly  from \eqref{eq:asympvarphim}.
\ep

\subsection{Proof of Theorem \ref{th:main}}
Let $\Gamma_h$ be defined by Lemma \ref{lem:compW} and let $\mathsf M_{V,h}$ be given by \eqref{eq:defMV}.
Notice that in the case $n_0\ge 3$, the matrix $\mathsf M_{V,h}$ has not to be diagonal and we vill use the following result. 
\begin{lemma}\label{le.GCM}  
 Introduce the symmetric positive semi-definite matrix 
$$\mathsf M_{V,h}^\Gamma=\Gamma_h^{-\frac 12}\mathsf M_{V,h}\Gamma_h^{-\frac 12}.$$
Then, there exists $\epsilon_0>0$ such that, for all $h$ small enough,
the  $n_0$ eigenvalues $ \beta_{1,h}\leq\dots \leq \beta_{n_{0},h}$ of 
$\mathsf M_{V,h}^\Gamma$  satisfy: $\beta_{1,h}=0$ and,
for all $k=2,\ldots, n_0$, $\beta_{k,h}\in\eee_{cl}(\mu_{k,h}^\Delta /\gamma_{k,h})$, where $\mu_{k,h}^\Delta$ is given by \eqref{EKW1}.
% $\mathsf M_{V,h}^\Gamma$ has $n_0$ eigenvalues $ (\beta_{k,h})_{k=1,\ldots,n_0}$ in the interval $[0,\epsilon_0 h]$, and one has 
%$\beta_{1,h}=0$ and 
%for all $k=2,\ldots, n_0$, $\beta_{k,h}\in\eee_{cl}(\mu_{k,h}^\Delta /\gamma_{k,h})$, where $\mu_{k,h}^\Delta$ is given by \eqref{EKW1}.
\end{lemma}
\noindent
\bp
First, observe that $\mathsf M_{V,h}$ admits $0$ as a simple eigenvalue (since $0$ is a simple eigenvalue of $\Delta_{V,h}$), so it is also a simple eigenvalue of $\mathsf M_{V,h}^\Gamma$. For  
$A=(a_{i,j})_{1\leq i,j\leq n_0}$ a matrix, we define $\tilde A:=(a_{i,j})_{2\leq i,j\leq n_0}$. We then have:
$$
\tilde{  \mathsf M}^\Gamma_{V,h}  =(\tilde \Gamma_h)^{-\frac 12}\tilde {\mathsf M}_{V,h}(\tilde \Gamma_h)^{-\frac 12}.
$$
Moreover, by~\eqref{a70} and~\eqref{a71}, the matrix $\tilde{  \mathsf M}_{V,h}$  writes   
$$
\tilde{  \mathsf M}_{V,h}=\Omega_h\big( \mathsf D_{V,h}+O(h^\infty)\big)\Omega_h,
$$
with $\Omega_h=\diag(e^{-S(\m_2)/h},\ldots,e^{-S(\m_{n_0})/h})$ and $\mathsf D_{V,h}=\diag(a_h(\m_2),\ldots,a_h(\m_{n_0}))$.  
Since $\tilde\Gamma_h$ is diagonal, one has 
$ \Omega_h\tilde\Gamma_h=\tilde\Gamma_h \Omega_h$ and  it follows that 
$\tilde{  \mathsf M}^\Gamma_{V,h}$ writes
$$
\tilde{  \mathsf M}^\Gamma_{V,h}=  \Omega_h  \big({\mathsf D}^\Gamma_{V,h} +O(h^\infty)\big) \Omega_h
$$
with ${\mathsf D}^\Gamma_{V,h}=\diag( \mu_{j,h}^\Delta/ \gamma_{j,h},j=2,\ldots,n_0)$. 
Hence $\tilde{  \mathsf M}^\Gamma_{V,h}$ admits a \textit{graded structure} in the sense of \cite[Definition A.1]{LePMi20}. We can thus apply \cite[Theorem A.4]{LePMi20} which yields the result.
\ep
\medskip

\noindent
We are now in position to prove Theorem \ref{th:main}. Let $U$ be a Morse function and assume that    \eqref{eq.H2} is satisfied.  Recall that we look for the singularities of  $E_{-+}^{-1}(\lambda)$ and that   $\lambda=0$ is always a singularity of $ E_{-+}^{-1}(\lambda)$ (since $\dim\Ker  (\mathsf P_h)=1$). Let us now look for the remaining singularities of $E_{-+}^{-1}(\lambda)$ in $ D(0, \epsilon_0 h) $. 
From  Lemma~\ref{le.GCM}, there exists a unitary change of basis $B_h$ such that $\mathsf M_{V,h}^\Gamma=B_h^*\diag( \beta_{1,h}, \ldots,\beta_{n_0,h})B_h$ (since $\mathsf M_{V,h}^\Gamma$ is symmetric and thus diagonalizable in an orthonormal basis). Combined with \eqref{eq:matEmp}, \eqref{eq:matEmp2}, and Lemma~\ref{lem:compW}, this yields
\be\label{eq:proofmain0}
-E_{-+}(\lambda)=\Gamma_h^{1/2}B_h^*\Big[\diag(\beta_{1,h}, \ldots,\beta_{n_0,h})-\lambda\mathsf I_{\C^{n_0}}-\mathsf R_h(\lambda)\Big]B_h\Gamma_h^{1/2},
\ee
where, for $h$ small enough,  $\mathsf R_h(\lambda)$ is holomorphic with respect to $\lambda\in D(0,\epsilon_0h)$ and 
\be\label{eq:proofmain1}
\mathsf R_h (\lambda)=O(\lambda^2h^{-\frac 32 }+\lambda h^{-\frac 12 } e^{-C/h}),
\ee
and where we have used that $\Gamma_h^{-1}=O(h^{-1/2})$ by~Lemma~\ref{lem:compW}. 
Hence, the singularities of $E_{-+}^{-1}(\lambda)$ in $ D(0, \epsilon_0 h) $ are exactly those of $L_h^{-1}$, where
\be\label{eq:proofmain2}
L_h(\lambda)=F_h(\lambda)-\mathsf R_h(\lambda), \text{ with } F_h(\lambda)=\diag(\beta_{1,h}-\lambda, \ldots,\beta_{n_0,h}-\lambda).
\ee 
Recall that $\beta_{1,h}=0$ and $\beta_{j,h}\in\eee_{cl}(\mu_{j,h}^\Delta/\gamma_{j,h})$ for all $j\geq 2$.  For all $j=2,\ldots,n_0$, using the asymptotic equivalents of $\mu_{j,h}^\Delta$ and $\gamma_{j,h}$ given in Theorem~\ref{th:vpWitten} and in Lemma~\ref{lem:compW},  one has when $h\to0$, 
using \eqref{eq.H1} which implies $\alpha(\m_j)>0$:
%if $\alpha(\m_j)>0$ (which is the case if~\eqref{eq.H1} holds) : 
\begin{equation}
\label{eq.DBeta}
\beta_{j,h}=\zeta_h ( \m_j) \,h\,e^{- \frac 2h S ( \m_j ) }  ,
\end{equation}
where  $\zeta_h(\m_j)\sim \sum_{k\geq 0}h^{\frac k 2}\zeta_k(\m_j)$ and $\zeta_0(\m_j)$  is given by~\eqref{e13}.
Let us now consider  $j\in\{2,\ldots,n_0\}$ and  $K\ge 2$. Denote $\ddd_{j,K}=D(\beta_{j,h},h^Ke^{-\frac 2h S(\m_j)})$ and let 
$\lambda\in\partial \ddd_{j,K}$. 
Since $S(\m_\ell)< S(\m_k)$ when $  \ell\ge k $,
for $h>0$ small enough,  the
${\ddd_{j,K}}$ are pairwise disjoint, their closures
are included in $D(0,\epsilon_0h)$, and for all $i\in \{2,\ldots,n_0\}$:  
\begin{equation}\label{eq.DBeta2}
\begin{split}
&\forall i>j,\;|\lambda-\beta_{i,h}|\geq |\beta_{j,h}-\beta_{i,h}|-|\lambda-\beta_{j,h}| \geq Che^{-\frac 2hS(\m_i) }\geq Che^{-\frac 2h S(\m_j)}\\
&\forall i<j,\;|\lambda-\beta_{i,h}|\geq |\beta_{j,h}-\beta_{i,h}|-|\lambda-\beta_{j,h}|  \geq Che^{-\frac 2h S(\m_j)}.  
\end{split}
\end{equation}
%where we have used that   $S(\m_\ell)< S(\m_k)$ if $  \ell\ge k $, and   that $K\ge 2$. 
Moreover, for $h$ small enough,
\begin{equation}\label{eq.DBeta3}
|\lambda|\ge Che^{-\frac 2h  S(\m_j)}.
\end{equation}
Consequently, for $\lambda\in\partial \ddd_{j,K}$, the matrix $F_h(\lambda)$ is invertible and 
$$F_h(\lambda)^{-1}=O(h^{-K}e^{\frac2hS(\m_j)}) \text{ on $\ddd_{j,K}$}.$$
 Combining this estimate with \eqref{eq:proofmain1} and reasoning as around~\eqref{eq.tour}, we prove that  for all $j=2,\ldots,n_0$, $K\geq 2$, and $h$ small enough, 
\begin{equation}\label{eq.i-fin}
\text{$L_h^{-1}(\lambda)$ (and thus $E_{-+}^{-1}(\lambda)$) admits a singularity $\alpha_{j,h,K}$ in $\ddd_{j,K}$,}
\end{equation}
so in particular $\alpha_{j,h,K}\neq \alpha_{i,h,K}$ when $i\neq j \in  \{2,\ldots,n_0\}$
(the ${\ddd_{j,K}}$ being pairwise disjoint).

In addition, since \eqref{eq.H1} holds, Theorem~\ref{th:1} implies that for all $h$ small enough, 
$\sigma(\mathsf  P_h)\cap\{ {\rm Re}(z)\le  \epsilon_0 h^2\}\setminus \{0\}$ is made of $n_0-1$ real eigenvalues
$0< \lambda_{2,h}\leq \ldots\leq \lambda_{n_0,h}$ (counted with algebraic multiplicity).
It then follows from \eqref{eq.start} that, for each $j=2,\dots,n_{0}$,
the eigenvalue $\lambda_{j,h}$
 satisfies
$\lambda_{j,h}=\alpha_{j,h,K}$ for all $K\ge 2$. 
Since $\alpha_{j,h,K}=\beta_{j,h} +O(h^{K}e^{-\frac 2h S(\m_j)})  $
for all $K\geq 2$ and $\beta_{j,h} \in \eee_{cl}(\mu_{j,h}^\Delta/\gamma_{j,h})$,
this 
completes the proof of Theorem~\ref{th:main}, using the asymptotic equivalent of $\mu_{j,h}^\Delta /\gamma_{j,h}$ as $h\to 0$ 
%for $j=2\ldots,n_0$ 
(see Theorem~\ref{th:vpWitten} and   Lemma~\ref{lem:compW}).

%with $\Re \alpha_{j,h}>0$ and $\alpha_{j,h}\neq \alpha_{i,h}$ if $i\neq j \in  \{2,\ldots,n_0\}$ (since they have different asymptotic equivalents as $h\to 0$ by injectivity of  $S$). 
%If~\eqref{eq.H1} holds, by Theorem~\ref{th:1}, $\sigma(\mathsf  P_h)\cap\{ {\rm Re}(z)\le  \epsilon_0 h^2\}\setminus \{0\}$ is composed of $n_0-1$ eigenvalues.   By~\eqref{eq.start} and since $\alpha_{j,h}\in \{ {\rm Re}(z)\le  \epsilon_0 h^2\}$ for $h$ small enough, for $h$ small enough, these $n_0-1$  eigenvalues are  $ \alpha_{2,h},\ldots, \alpha_{n_0,h}$. 
%This completes the proof of Theorem~\ref{th:main} since $\alpha_{j,h}=\beta_{j,h} +O(h^{K}e^{-\frac 2h S(\m_j)})  $, $\beta_{j,h} \in \eee_{cl}(\mu_{j,h}^\Delta/\gamma_{j,h})$, and by the asymptotic equivalents of $\mu_{j,h}^\Delta /\gamma_{j,h}$ as $h\to 0$ for $j=2\ldots,n_0$ (see Theorem~\ref{th:vpWitten} and   Lemma~\ref{lem:compW}).

%%%%%%

\subsection{The case when~\eqref{eq.H1} is not satisfied}
 In this section, we prove Theorem~\ref{th:alpha}, where, compared to Theorem~\ref{th:main},  we do no longer assume that \eqref{eq.H1} holds but that \eqref{eq.H1-alpha=0} holds. We can thus no longer use Theorem~\ref{th:1}
 as we did 
 at the very end of the proof of Theorem~\ref{th:main} to  
 say that  $\mathsf P_h$ admits $n_0-1$ nonzero eigenvalues (counted with multiplicity) in   $\{ {\rm Re}(z)\le  \epsilon_0 h^2\}$.
% conclude that the $n_0-1$ nonzero eigenvalues of $\mathsf P_h$ in   $\{ {\rm Re}(z)\le  \epsilon_0 h^2\}$ (counted with multiplicity) were precisely the singularities  $\alpha_{j,h,K}$, $j=2,\ldots,n_0$, of $\lambda \mapsto E_{-+}^{-1}(\lambda)$ (actually independent of $K\geq 2$).
%The assumption~\eqref{eq.H1} has been made so that we could use Theorem~\ref{th:1} which  has only been used at the very end of the proof of Theorem~\ref{th:main} to  conclude that the $n_0-1$ nonzero eigenvalues of $\mathsf P_h$ in   $\{ {\rm Re}(z)\le  \epsilon_0 h^2\}$ are the singularities  $\{\alpha_{j,h}, j=2,\ldots,n_0\}$ of $\lambda \mapsto E_{-+}^{-1}(\lambda)$. 
%Without~\eqref{eq.H1}, 
%we cannot use Theorem~\ref{th:1}, but 
We can  however make use of all the intermediate results in the proof of Theorem~\ref{th:main} until \eqref{eq.i-fin} (included), except that one has the following minor scaling changes:
\begin{enumerate}
\item Equation \eqref{eq.DBeta} must be changed into 
\begin{equation}\label{eq.dLalpha}
\beta_{j,h}=\zeta_h ( \m_j)\,\sqrt h\, e^{- \frac 2h S ( \m_j ) }  ,
\end{equation}
 where 
$ \zeta_h(\m_j)\sim \sum_{k\geq 0}h^{\frac k 2} \zeta_k(\m_j)$ with $  \zeta_0 ( \m_j) $ satisfying~\eqref{e-alpha2}. 
\item In Equations  \eqref{eq.DBeta2} and  \eqref{eq.DBeta3}, all the $Ch$  must be replaced by $C\sqrt h$.  
\end{enumerate}
Let $j\in\{2,\dots,n_{0}\}$ and note that, for $h$ small enough, the sequence $(\alpha_{j,h,K})_{K\geq 2}$
defined in the proof of  Theorem~\ref{th:main}
is stationary by analyticity of the nontrivial map $\lambda\mapsto E_{-+}(\lambda)$. 
We denote by $\alpha_{j,h}$ its limit
and recall that, according to \eqref{eq.start}, $\{0, \alpha_{2,h},\ldots, \alpha_{n_0,h}\}\subset \sigma(\mathsf P_h)$. 
In addition, the $\alpha_{j,h}$'s  are exponentially small and thus belong to  $D(0,ch^2)$
for any $c>0$  and all $h$ small enough. Let us now prove that, for $h$ small enough, 
$0$ and $\alpha_{j,h}$, $j=2,\ldots,n_0$,
    are the only singularities of $\lambda \mapsto E_{-+}^{-1}(\lambda)$   in  $D(0,ch^2)$, that they are all real, 
  and have
    geometric multiplicity $1$ as eigenvalues of $\mathsf P_h$.  
Recall that by~\eqref{eq:proofmain0}, $\lambda\in D(0,\epsilon_0h)$  is a singularity of $E_{-+}^{-1}(\lambda)$ if and only if it is a singularity of $L_h(\lambda)^{-1}$ (see~\eqref{eq:proofmain2}). Let us denote, for $\lambda\in \mathbb C$,  
$$d_h(\lambda)= \det L_h(\lambda)=\det (F_h(\lambda)-\mathsf R_h(\lambda)),$$
which is holomorphic on $D(0,\epsilon_0h)$. 
For all $\lambda \in \partial D(0,ch^2)$, using~\eqref{eq:proofmain1}, we get  
$\mathsf R_h(\lambda)=O(h^4h^{-3/2})=O(h^{5/2})$,
and by~\eqref{eq.dLalpha}, 
$$\big |\det F_h(\lambda)\big |=\big |\lambda\times \Pi_{j=2}^{n_0}(\lambda-\beta_{j,h})\big|= c^{n_0}h^{2n_0}(1+o_h(1)).$$
Thus,  for $h$ small enough, it holds  uniformly on $\partial D(0,ch^2)$:
\begin{equation}\label{eq.Dh0}
d_h(\lambda)=\lambda\times \Pi_{j=2}^{n_0}(\lambda-\beta_{j,h}) +O(h^{5/2}h^{2(n_0-1)})=\det F_h(\lambda) (1+o_h(1)).
\end{equation}
In particular, for any $\lambda\in\partial D(0,ch^2)$, one has 
$$
|d_h(\lambda)-\det F_h(\lambda)|<|\det F_h(\lambda)|,
$$
which implies by Rouch\'e's theorem that $d_h(\lambda)$ and $\det F_h(\lambda)$ have the same number $n_{0}$ of zeros (counted with multiplicity)
in $D(0,ch^2)$.
These $n_0$ zeros are thus $0$ and $\alpha_{j,h}$, $j=2,\ldots,n_0$.  Let us recall that
the $\alpha_{j,h}$ are pairwise disctinct and satisfy
 $\alpha_{j,h}= \beta_{j,h}+O(h^Ke^{-\frac 2h S(\m_j)})$ for all $K\ge 2$ and $j=2,\ldots,n_0$.
% Then, by \eqref{eq.dLalpha},  $0$ and $\alpha_{j,h}$, for $j=\{2,\ldots,n_0\}$, have  different asymptotic equivalents as $h\to 0$ (because $S$ is injective, by~\eqref{eq.H2}), and thus
 They are then all simple zeros of $d_h(\lambda)$ (and thus $\dim \Ker E_{-+}(\alpha_{j,h})=1$ for $j=2,\ldots,n_0$). Using also \eqref{eq.THbis} and 
 \cite[Equation (2.7)]{SjZw07}, $\dim \Ker (\mathsf P_h-\alpha_{j,h})=\dim \Ker \mathsf T_h(\alpha_{j,h})=\dim \Ker E_{-+}(\alpha_{j,h})=1$, for all $j=\{2,\ldots,n_0\}$. 
Lastly, since the operator $\mathsf P_h$ has real coefficients, 
its spectrum is stable by complex conjugation. Hence, since moreover $0$ and the $\alpha_{j,h}$, $j=2,\ldots,n_0$, have  different asymptotic equivalents as $h\to 0$, one has $\overline{\alpha_{j,h}}={\alpha_{j,h}}$ for every  $j=2,\ldots,n_0$. 
To conclude the proof of Theorem~\ref{th:alpha}, it then just remains to 
use that, for every  $j=2,\ldots,n_0$, $\beta_{j,h} \in \eee_{cl}(\mu_{j,h}^\Delta/\gamma_{j,h})$,
and the asymptotic equivalents of $\mu_{j,h}^\Delta /\gamma_{j,h}$ as $h\to 0$ 
%for $j=2\ldots,n_0$ 
given by Theorem~\ref{th:vpWitten} and   Lemma~\ref{lem:compW}.
\bigskip

\noindent
\textbf{Acknowledgement}\\
This work was  supported by the ANR-19-CE40-0010, Analyse Quantitative de Processus M\'etastables (QuAMProcs).
B.N. is supported by the grant IA20Nectoux from the Projet I-SITE Clermont CAP 20-25. The authors are grateful to Pierre Monmarch\'e for fruitful discussions.

\bibliographystyle{amsplain}
\bibliography{ref_zigzag}
\end{document}